\documentclass[12pt]{article}
\usepackage{amsfonts, amsmath, amssymb,epsfig}
\usepackage[latin1]{inputenc}
\usepackage[english]{babel}
\usepackage{color}
\newtheorem{thm}{Th\'eor\`eme}[section]
\newtheorem{cor}[thm]{Corollaire}
\newtheorem{lem}[thm]{Lemme}
\newtheorem{pro}[thm]{Proposition}
\newtheorem{dfn}[thm]{D\'efinition}
\newtheorem{rmk}[thm]{Remark}
\newtheorem{expl}[thm]{Exemple}

\def\dessous#1\sous#2{\mathrel{\mathop{\kern0pt#2}\limits_{#1}}}

\newcommand{\R}{\mathbb R}
\newcommand{\C}{\mathbb C}

\newcommand{\1}{1 \! \! {\rm I}}

\newcommand{\beq}{\begin{eqnarray}}
\newcommand{\eeq}{\end{eqnarray}}
\newcommand{\bpro}{\begin{pro}}
\newcommand{\epro}{\end{pro}}
\newcommand{\blem}{\begin{lem}}
\newcommand{\elem}{\end{lem}}
\newcommand{\bdfn}{\begin{dfn}}
\newcommand{\edfn}{\end{dfn}}
\newcommand{\bcor}{\begin{cor}}
\newcommand{\ecor}{\end{cor}}
\newcommand{\bthm}{\begin{thm}}
\newcommand{\ethm}{\end{thm}}
\newcommand{\bex}{\begin{expl}}
\newcommand{\eex}{\end{expl}}
\newcommand{\brmk}{\begin{rmk}}
\newcommand{\ermk}{\end{rmk}}
\newcommand{\benum}{\begin{enumerate}}
\newcommand{\eenum}{\end{enumerate}}
\newcommand{\bitem}{\begin{itemize}}
\newcommand{\eitem}{\end{itemize}}

\begin{document}

\begin{titlepage}
\begin{flushright}
ICMPA-MPA/2012/20\\
\end{flushright}
   \begin{center}
     {\ }\vspace{0.5cm}
      {\bf {On nonlinear coherent states properties for electron-phonon dynamics}}

 \vspace{0.5cm}

Isiaka Aremua$^{ a}$, Mahouton Norbert Hounkonnou$^\dag$ and Ezinvi Balo\"itcha$^{\dag\dag}$

\vspace{1cm}

%   \vspace{1cm}

       {\em  International Chair of Mathematical Physics
and Applications} \\
{\em ICMPA-UNESCO Chair}\\
{\em University of Abomey-Calavi}\\
{\em 072 B.P. 50 Cotonou, Republic of Benin}\\

{\em E-mail: {\tt  claudisak@yahoo.fr$^a$, norbert.hounkonnou@cipma.uac.bj$^\dag$\footnote{Corresponding author
 (with copy to hounkonnou@yahoo.fr)},  ezinvi.baloitcha@cipma.uac.bj$^{\dag\dag}$
}}

\vspace{0.5cm}

    \vspace{2.0cm}
           \today
 %        \clearpage
          \begin{abstract}
\noindent
This work addresses a  construction  of a  dual pair of nonlinear coherent states (NCS) in the context of 
changes of bases in the underlying Hilbert space for a model pertaining to  the condensed 
matter physics, which obeys a $f$-deformed Heisenberg algebra. 
The existence and properties of reproducing kernel  in the NCS Hilbert space are studied and discussed;    the  probability density
and its dynamics  in the  basis of constructed coherent states are provided.   
A  Glauber-Sudarshan $P$-representation of the density matrix   and  relevant issues  related to the reproducing kernel 
 properties are presented. Moreover,  a  NCS  quantization 
 of classical phase space observables is performed and illustrated in a concrete example of $q$-deformed coherent states. 
Finally,  an exposition of quantum optical  properties is given.

\end{abstract}

\end{center}

\end{titlepage}

\section{Introduction}

Coherent states (CS), known as the closest states to classical behaviour of a system, play an important role in   theoretical and 
experimental
 physics.
First introduced as venerable objects by Schr\"{o}dinger \cite{sch} in 1926 for the harmonic oscillator potential, 
they are since  at the core of   research directions    in quantum optics,  which is an ideal testing ground for ideas of 
quantum theory. Much work has been done through their theoretical generalizations including their experimental 
generations and applications. Then, the generalization based on group symmetry approach has led to define CS for arbitrary  
Lie algebras such as $su(1,1)$, $su(2)$ which have found numerous applications in quantum optics \cite{perelomov}, \cite{zhang}. 

Further generalizations also extend to so-called  
nonlinear CS (NCS) or $f$-CS \cite{man'ko} induced by nonlinear algebras  referring to  
 $f$-deformed oscillator algebras initially introduced by 
Jannussis {\it{ et al.}} \cite{jannussis-brodimas} and  Man'ko {\it{et al}}. \cite{man'ko},  and   spanned by the  ordinary Fock-Heisenberg  
generators $\{a,a^{\dag},N\}$ coupled to a free continuous regular function $f$ of the number operator $N, $ such that 
\beq
A^{-} &=& af(N), \qquad  A^{\dag} = f(N)a^{\dag} \cr
{N} &=& A^{\dag}A^{-} = Nf^{2}(N), \qquad [A^{-},A^{\dag}] = \{N+1\} - \{N\}.
\eeq
The  $f$-deformed quantum algebras offer the advantage to be well represented in the ordinary Fock-Hilbert space. Although they are not merely mathematical objects, 
NCS are useful to analyse many  quantum mechanical systems. For instance, in \cite{matos}, it  has been shown that they are important in the description of the 
motion of a trapped ion and 
 have some ``non-classical properties'' such as squeezing, amplitude-squared squeezing, antibuching,  
sub-Poissonian behavior.    

 NCS construction uses  an operator similar to the displacement operator, 
 generally denoted by $D(z), z \in \C$ with  \cite{roy-roy}
\beq
D(z) = e^{za^{\dag} - \bar{z}a}, \,\;\; [a, a^{\dag}] = \1
\eeq
such that the CS parametrized by $z$ are given by the action of $D(z)$ on the ground state $|0\rangle:$
\beq
|z\rangle= D(z)|0\rangle.
\eeq
 This  is in connection with the method proposed in \cite{ali-roknizadeh-tavassoly}
to define NCS  by changing  bases in the underlying Hilbert space, involving   an interesting duality between pairs of 
constructed generalized CS.

Another class of CS describing quantum optical models  encompasses  vector coherent states (VCS)    (\cite{berube-hussin-nieto}, \cite{daoud-hussin}, \cite{ali-englis-gazeau}),
  used for example in the 
study of spectra of two-level atomic systems placed in electromagnetic fields like the Jaynes-Cummings model \cite{jcum}.
For more details, see  \cite{joben} (and references therein)  presenting a  formulation of VCS for nonlinear spin-orbit Hamiltonian model
in terms of 
the 
matrix eigenvalue problem for generalized annihilation operators.
 A formalism of VCS construction for a system of $M$ Fermi-type modes associated with $N$
bosonic modes \cite{hong}-\cite{yang} 
 was also given  in \cite{a-hk}. The defined VCS satisfy, in this context, required mathematical properties of continuity, resolution 
of identity, temporal stability and action identity.

This work, generalizing    previous investigations  (see \cite{a-hk} and references therein), develops  a construction  of a  dual pair of NCS, 
based on   a $f$-deformed algebra,  to describe such phenomena like the electron-phonon dynamics in condensed 
matter physics.

The paper is organized as follows. In Section 2,  the physical model and the formalism of
 nonlinear coherent states (NCS) construction are described.  Section 3 deals with  the  probability density
and its dynamics  in the  basis of constructed coherent states.   The  existence and properties of reproducing kernel  in the NCS Hilbert
 space are studied and discussed in Section 4. 
In Section 5, the  Glauber-Sudarshan $P$-representation of the density matrix  is elaborated in both
the NCS and nonlinear VCS (NVCS);   
the associated  reproducing kernel properties are studied. 
Section 6  presents   the NCS quantization of the  complex plane. In Section 7, relevant quantum optical properties are analyzed.
 The last section is devoted to concluding remarks.

\section{CS for electron-phonon dynamics}
In this section, we  construct the  CS for a  physical model describing the electron-phonon dynamics given by the following Hamiltonian 
 (with $\hbar =
1$) \cite{hong}-\cite{a-hk}:
\beq{\label{hamil} } 
H = \sum_{i=1}^{N_B}\omega_i a^{\dag}_i
a_i + \sum_{j=1}^{M}\epsilon_{j}c^{\dag}_jc_j +
\sum_{i=1}^{N_B}\sum_{l=1}^{M}g_l c^{\dag}_l c_l (a^{\dag}_i + a_i)
\eeq 
where the following commutation rules hold:
\beq{\label{commuta}}
 &&  [a_{_{i}}, a^{^{\dag}}_{k}] = \delta_{ik}\mathbb I \qquad
\{c^{\dag}_{j}, c_{l}\} = \delta_{jl}\mathbb I \cr
&&  [a_{i}, a_{k}] = 0 \qquad
[a^{\dag}_{i}, a^{\dag}_{k}]=0 \qquad   \{c_{j}, c_{l}\} = 0 \qquad
\{c^{\dag}_{j}, c^{\dag}_{l}\}=0 \qquad    \eeq
with $1\leq i, k \leq N_B,$  $1 \leq j, l \leq
M$. The set  $\{a_{i},a^{\dag}_{i}, \1\}$
for each $i \, (1\leq i \leq N_B)$ spans the ordinary Fock-Heisenberg
oscillator algebra. A similar model,  describing an
interaction between a single mode, $(a, a^\dag)$, of the radiation
field with two Fermi type modes, was also
investigated by Simon and Geller who outlined  some physical
aspects of the ensemble-averaged excited-state population dynamics
 \cite{simon-geller} and showed its   relevance in the
study of electron-phonon dynamics in an ensemble of nearly isolated
nanoparticles, in the context of quantum
effects in condensed matter systems. The vibrational spectrum of a
nanoparticle is here provided by the localized electronic impurity
states in doped nanocrystal
  \cite{hong, yang}. The impurity states are used to probe the
  energy relaxation by phonon emission.

The  deformed version of the  Hamiltonian (\ref{hamil}) is obtained  by performing the correspondences  
\beq{\label{eq1}} 
a_l \rightarrow  a_lf_{l}(N_{l}) \qquad a^{\dag}_{l} \rightarrow f_{l}(N_{l})a^{\dag}_l \qquad 1\leq l \leq N_B
\eeq
with  the following  nonlinear commutation rules for the bosonic operators:
\beq{\label{eq2}}
[N_{l},a_{l}f_{l}(N_{l})] = -a_{l}f_{l}(N_{l}), \quad [N_{l},f_{l}(N_{l})a^{\dag}_{l}]=f_{l}(N_{l})a^{\dag}_{l}, \cr
[a_{l}f_{l}(N_{l}),f_{l}(N_{l})a^{\dag}_{l}]=(N_{l}+1)f_{l}^{2}(N_{l}+1)-N_{l}f_{l}^{2}(N_{l}),
\eeq

  $f_{l}(N_{l})$ being a reasonably well behaved real function of the number 
operator $N_{l}.$  
%The nonlinear algebra depends on the choice of the  \textquotedblleft nonlinearity \textquotedblright 
%function $f_{l}(N_{l})$.
 If
$f_{l}(N_{l}) = 1,$ then the nonlinear algebra in (\ref{eq2}) reduces to the ordinary non-deformed oscillator algebra.

Consider the following self-adjoint operator    given by 
\beq{\label{eq9}} 
B_{[{\bf k}]_{l}}=\omega_{l}A^{\dag}_{[{\bf k}]_{l}}A_{[{\bf k}]_{l}}+\frac{\epsilon_{[\bf k]}}{N_B}-\frac{g^{2}_{[\bf k]}}{\omega_{l}} \quad 
\mbox{with} 
\quad A_{[{\bf k}]_{l}} = a_{l}f_{l}(N_{l})+\frac{g_{[\bf k]}}{\omega_{l}} 
\eeq 
 
where $g_{[\bf k]}, \epsilon_{[\bf k]}$ are defined as  \cite{a-hk}

\beq g_{[\bf k]}:=\sum_{k_{j} \in[{\bf
k}] }k_{j}g_{j}\qquad \epsilon_{[{\bf k}]}:=\sum_{k_{j} \in[{\bf
k}] }k_{j}\epsilon_{j} \qquad 1\leq j \leq M  \qquad [{\bf k}] \in \Gamma,
\eeq
 
\beq{\label{set00}}
\Gamma = \{(0,0,\cdots, 0,0), (1,0, \cdots, 0,0), \dots, (1,1\cdots, 1,1) \}
\eeq

 and the operator  

\beq{\label{eq12}}
B_{[{\bf k}]_{[\bf n]}}=\sum_{l=1}^{N_B}\omega_{l}A^{\dag}_{[{\bf k}]_{l}}A_{[{\bf k}]_{l}}+ 
\epsilon_{[\bf k]}-g^{2}_{[\bf k]}\sum_{l=1}^{N_B}\frac{1}{\omega_{l}} \quad \mbox{with} \mbox \quad  [{\bf{n}}]: = n_1, n_2, \dots,  n_{N_B}.
\eeq

In the next subsection, we define
the  operators useful for our construction by transforming CS into their deformed correspondents with an
operator  $T,$ and conversely into their duals with the inverse operator $T^{-1}$, such that $TT^{-1} = \mathbb I,$ the
identity operator of the considered  Hilbert space.

\subsection{Rescaled basis states, eigenvalues and eigenvectors}
For all $l$, $1\leq l \leq N_B,$  ${\mathfrak H}_{l}$ denotes the separable  Hilbert space  spanned by the 
eigenvectors  $\phi_{n_{l}},n_{l}=0,1,2,\dots ,\infty$ of the number operator $N_{l}.$

On  ${\mathfrak H}_{l},$ let  $T$  be an operator densely defined and closed in the  domain   denoted by  
${\mathcal D}(T).$ Suppose that  $T^{-1}$ 
exists and is densely defined  with domain ${\mathcal D}(T^{-1})$. Moreover, the vectors 
$\phi_{n_{l}} \in \mathcal D(T) \cap \mathcal D(T^{-1})$ for all $n_{l}$ and there exist non-empty open sets $\mathcal D_{T}$ 
and $\mathcal D_{T^{-1}}$ in $\C$ such that $\eta_{z_{l}} \in \mathcal D(T),   \forall z_{l} \in \mathcal D_{T}$
 and $\eta_{z_{l}} \in \mathcal D(T^{-1}),  \forall z_{l} \in \mathcal D_{T^{-1}}$.

The operator  $T$ being densely defined and closed, then the operator $\mathcal S$ such that $\mathcal S = T^{*}T,$  with domain 
$\mathcal D(\mathcal S)$,  where $T^{*}$ is 
the operator adjoint of $T,$
is self-adjoint.

Let  ${\mathfrak H}^{\mathcal S}_{l}$ be the completion of  $ {\mathcal D}(T)$ in the scalar product
\beq{\label{eq13}} 
\langle \phi_{n_{l}}|\psi_{n_{l}}\rangle_{{\mathfrak H}^{\mathcal S}_{l}} 
= \langle \phi_{n_{l}}|T^{*}T \psi_{n_{l}}\rangle_{{\mathfrak H}_{l}}=\langle \phi_{n_{l}}| \mathcal S\psi_{n_{l}}\rangle_{{\mathfrak H}_{l}}, 
 \quad |\psi_{n_{l}}\rangle \in \mathcal D(\mathcal S), 
\eeq 
where ${\mathfrak H}^{\mathcal S}_{l}$ is spanned by the vectors
\beq{\label{eq14}}
 |\phi^{\mathcal S}_{n_{l}}\rangle= T^{-1}|\phi_{n_{l}}\rangle,   \quad |\phi_{n_{l}}\rangle \in \mathcal D({T^{-1}}), 
\eeq 
and ${\mathfrak H}^{\mathcal S^{-1}}_{l}$ the completion of ${\mathcal D}({T^{*}}^{-1})$  in the scalar product
\beq{\label{eq15}} 
\langle \phi_{n_{l}}| \psi_{n_{l}}\rangle_{{\mathfrak H}^{\mathcal S^{-1}}_{l}} = \langle \phi_{n_{l}}|T^{-1}{T^{*}}^{-1}
\psi_{n_{l}}\rangle_{{\mathfrak H}_{l}} = \langle \phi_{n_{l}}|\mathcal S^{-1} \psi_{n_{l}}\rangle_{{\mathfrak H}_{l}}, 
  \quad |\psi_{n_{l}}\rangle \in \mathcal D(\mathcal S^{-1}), 
\eeq
 where   
$\mathcal D(\mathcal S^{-1})$ is the    domain of $\mathcal S^{-1}$, with ${\mathfrak H}^{\mathcal S^{-1}}_{l}$  spanned by the vectors
\beq{\label{eq16}} 
|\phi^{\mathcal S^{-1}}_{n_{l}}\rangle=T|\phi_{n_{l}}\rangle,  \quad |\phi_{n_{l}}\rangle \in \mathcal D({T}).
\eeq 

Moreover, if the spectrum of $\mathcal S$ is bounded away from zero, then $\mathcal S^{-1}$  is bounded and there  follow the inclusions 
\beq
\mathfrak H^{\mathcal S}_{l} \subset \mathfrak H_{l}  \subset \mathfrak H^{\mathcal S^{-1}}_{l}.
\eeq
The so-defined Hilbert spaces $\mathfrak H^{\mathcal S}_{l}, \mathfrak H_{l}$ and $\mathfrak H^{\mathcal S^{-1}}_{l}$ are called a
 {\it Gelfand triple}.
 
 On  ${\mathfrak H}^{\mathcal S}_{l}$ the transformed of the operators  
$a_{l},a^{\dag}_{l},N_{l}$ defined on ${\mathfrak H}_{l}$ are \cite{ali-roknizadeh-tavassoly}
\beq{\label{eq17}} 
a^{\mathcal S}_{l}=T^{-1}a_{l}T,\qquad {a^{\mathcal S}_{l}}^{\dag} =T^{-1}a^{\dag}_{l}T, \qquad  N^{\mathcal S}_{l}=T^{-1}N_{l}T,
\eeq
and their actions on ${\mathfrak H}^{\mathcal S}_{l}$ defined by
\beq{\label{eq18}} 
a^{\mathcal S}_{l}|\phi^{\mathcal S}_{n_{l}} \rangle:=\sqrt{n_{l}}| \phi^{\mathcal S}_{n_{l}-1}\rangle, \quad 
{a^{\mathcal S}_{l}}^{\dag}|\phi^{\mathcal S}_{n_{l}}\rangle:=\sqrt{n_{l}
+1}|\phi^{\mathcal S}_{n_{l}+1}\rangle,\quad N^{\mathcal S}_{l}|\phi^{\mathcal S}_{n_{l}}\rangle:=n_{l}
|\phi^{\mathcal S}_{n_{l}}\rangle.
\eeq 
These operators satisfy the same commutation relations as $a_{l}, a^{\dag}_{l}$ and $N_{l}$:
\beq{\label{eq19}} 
[a^{\mathcal S}_{l},{a^{\mathcal S}_{l}}^{\dag}]
=\mathbb I,\quad [a^{\mathcal S}_{l},N^{\mathcal S}_{l}]=a^{\mathcal S}_{l},\quad [{a^{\mathcal S}_{l}}^{\dag},N^{\mathcal S}_{l}]
=-{a^{\mathcal S}_{l}}^{\dag}.
\eeq 
Also the operators $a^{\mathcal S^{-1}}_{l}, {a^{\mathcal S^{-1}}_{l}}^{\dag}, N^{\mathcal S^{-1}}_{l}$ are defined on 
${\mathfrak H}^{\mathcal S^{-1}}_{l}$
and their actions on ${\mathfrak H}^{\mathcal S^{-1}}_{l}$ are given in an analogous manner as in (\ref{eq18}); 
 they satisfy the commutation relations 
\beq{\label{eq20}} 
[a^{\mathcal S^{-1}}_{l},{a^{\mathcal S^{-1}}_{l}}^{\dag}]=\mathbb I, \quad [a^{\mathcal S^{-1}}_{l},N^{\mathcal S^{-1}}_{l}]=
a^{\mathcal S^{-1}}_{l},\quad [{a^{\mathcal S^{-1}}_{l}}^{\dag},
N^{\mathcal S^{-1}}_{l}]=-{a^{\mathcal S^{-1}}_{l}}^{\dag}.
\eeq
The operators 
$T,T^{-1},\mathcal S$ can be  written as follows:  
\beq{\label{eq21}} 
&&T:=T(N_{l})=\sum_{n_{l}=0}^{\infty}t(n_{l})|\phi_{n_{l}}\rangle \langle \phi_{n_{l}}|,\quad 
T^{-1}:=T(N_{l})^{-1}=\sum_{n_{l}=0}^{\infty}\frac{1}{t(n_{l})}|\phi_{n_{l}}\rangle \langle \phi_{n_{l}}|,\cr
&&\mathcal S:=\mathcal S(N_{l})=\sum_{n_{l}=0}^{\infty}t(n_{l})^{2}|\phi_{n_{l}}\rangle \langle \phi_{n_{l}}|.
\eeq
For all $l$, the action of  $f_{l}(N_{l})$  on the basis vectors $|\phi_{n_{l}}\rangle$ of the separable   Hilbert space 
${\mathfrak H}_{l}$ can be also defined as:
\beq{\label{eq22}}
f_{l}(N_{l})|\phi_{n_{l}}\rangle:=\frac{t(n_{l})}{t(n_{l}-1)}|\phi_{n_{l}}\rangle=f_{l}(n_{l})|\phi_{n_{l}}\rangle,
\eeq
where 
\beq{\label{eq23}}
t(n_{l})=f_{l}(n_{l})f_{l}(n_{l}-1)\cdots f(1):=f_{l}(n_{l}) !
\eeq 
%where the  $t(n_{l})$
 are real numbers, possessing the properties:
\bitem
\item $t(0)=1$ and $t(n_{l})=t(n'_{l})$ if and only if $n_{l}=n'_{l};$
\item $ 0 < t(n_{l}) < \infty$;
\item the Cauchy criterium  implies that the finiteness condition for the limit  
\beq
\lim\limits_{n_{l} \to \infty}\left[\frac{t(n_{l})}{t(n_{l}+1)}\right]^{2}\frac{1}{n_{l}+1}=\rho < \infty,
\eeq
holds. This condition implies that the series  
$S(r^{2}):=\sum_{n_{l}=0}^{\infty}\frac{r^{2n_{l}}}{[t(n_{l})]^{2}n_{l} !}$ converges for all $r<L=1/\sqrt{\rho}.$
\eitem
For all $l$, we define the function of the operator $N_{l}, f_{l}(N_{l})^{-1}$
acting on the basis vectors  $|\phi_{n_{l}}\rangle$ of  ${\mathfrak H}_{l}$ in the following manner:
\beq{\label{eq24}} 
f_{l}(N_{l})^{-1}|\phi_{n_{l}}\rangle:= \frac{t(n_{l}-1)}{t(n_{l})}|\phi_{n_{l}}\rangle=\frac{1}{f_{l}(n_{l}) }|\phi_{n_{l}}\rangle.
\eeq 
The real numbers $t(n_{l})$ satisfying the   above two first  properties are such  that the finiteness condition for the limit
\beq
\lim\limits_{n_{l} \to \infty}\left[\frac{t(n_{l}+1)}{t(n_{l})}\right]^{2}\frac{1}{n_{l}+1}=\tilde{\rho}
<\infty,
\eeq 
implies that the series  $S'(r^{2}):=\sum_{n_{l}=0}^{\infty}\frac{r^{2n_{l}}(t(n_{l}))^{2}}{n_{l} !}$ converges for all 
$r<\tilde{L}=1/\sqrt{\tilde{\rho}}.$
Since $|\phi^{\mathcal S^{-1}}_{n_{l}}\rangle = T |\phi_{n_{l}}\rangle,$ then
\beq {\label{eq25}}
|\phi^{\mathcal S^{-1}}_{n_{l}}\rangle = t(n_{l})|\phi_{n_{l}}\rangle = f_{l}(n_{l}) !|\phi_{n_{l}}\rangle.
\eeq
Let us  set $A_{l}=a^{\mathcal S}_{l}$ when $a^{\mathcal S}_{l}$ acts on ${\mathfrak H}_{l}$, and  
$A'_{l}=a^{\mathcal S^{-1}}_{l}$ when $a^{\mathcal S^{-1}}_{l}$ acts on ${\mathfrak H}_{l}.$
Since  $T,T^{-1}$ and $\mathcal S$ are positive operators,   the adjoints of  $A_{l}$ and $A'_{l}$ on ${\mathfrak H}_{l}$
take  the form:
\beq{\label{eq26}}
A^{\dag}_{l}=Ta^{\dag}_{l}T^{-1}= {a^{\mathcal S^{-1}}_{l}}^{\dag},\qquad A'^{\dag}_{l}=T^{-1}a^{\dag}_{l}T={a^{\mathcal S}_{l}}^{\dag}.
\eeq
From the relations 
\beq{\label{rec1}} 
f_{l}(N_{l})|\phi_{n_{l}}\rangle= f_{l}(n_{l})|\phi_{n_{l}}\rangle, \, |\phi^{\mathcal S}_{n_{l}}\rangle = 
\frac{1}{f_{l}(n_{l}) !}|\phi_{n_{l}}\rangle
\eeq 
and
\beq{\label{rec2}} 
f_{l}(N_{l})^{-1}|\phi_{n_{l}}\rangle = \frac{1}{f_{l}(n_{l})}|\phi_{n_{l}}\rangle, \, 
|\phi^{\mathcal S^{-1}}_{n_{l}}\rangle = f_{l}(n_{l}) !|\phi_{n_{l}}\rangle,
\eeq 
we infer their   actions  as 
\beq\label{oprealiz00} 
A_{l}|\phi_{n_{l}}\rangle = \sqrt{n_{l}} f_{l}(n_{l})|\phi_{n_{l}-1}\rangle, \qquad  
A^{\dag}_{l}|\phi_{n_{l}}\rangle =\sqrt{n_{l}+1} f_l(n_{l}+1)|\phi_{n_{l}+1}\rangle.
\eeq 
The vectors $|\phi^{\mathcal S}_{n_{l}}\rangle$ and $|\phi^{\mathcal S^{-1}}_{n_{l}}\rangle,$ 
defined in (\ref{rec1}) and (\ref{rec2}),  are called {\it rescaled basis states} \cite{ali-roknizadeh-tavassoly}.

Then we get
\beq\label{oprealiz01}
A_{l}=a_{l}f_{l}(N_{l}),\,\,  A^{\dag}_{l}=f_{l}(N_{l})a^{\dag}_{l}, \,\, 
 A'_{l}=a_{l}f_{l}(N_{l})^{-1}, \,\, A'^{\dag}_{l}=f_{l}(N_{l})^{-1}a^{\dag}_{l}
\eeq

with the following relations  
\beq\label{oprealiz02} 
[A_{l},A'^{\dag}_{l}] = \mathbb I \qquad 
[A'_{l},A^{\dag}_{l}]  = \mathbb I  \qquad  1 \leq l \leq N_B.
\eeq

From the relations (\ref{oprealiz00}),   the operators $A_{l},  A^{\dag}_{l}$ and $N_l$  can be expressed as

\beq\label{opexp}
A_{l} &=& \sum_{n_{l} = 0}^{\infty} \sqrt{n_{l}} f_{l}(n_{l})|\phi_{n_{l}-1}\rangle \langle \phi_{n_{l}}|, \qquad  
A^{\dag}_{l} = \sum_{n_{l} = 0}^{\infty} \sqrt{n_{l}} f(n_{l})|\phi_{n_{l}}\rangle \langle \phi_{n_{l}-1}|  \cr
N_l &=& \sum_{n_{l} = 0}^{\infty} 
n_{l}|\phi_{n_{l}}\rangle  \langle \phi_{n_{l}}|. 
\eeq

For all  $l$, we define, on the  Hilbert space  ${\mathfrak H}_{l},$ the momentum operator $P_{l}$ as
\beq\label{moment00} 
P_{l} := \frac{a_{l}f_{l}(N_{l}) - f_{l}(N_{l})^{-1}a^{\dag}_{l}}{i \sqrt{2}} \qquad \mbox{with} \qquad 
[a_{l}f_{l}(N_{l}),\sqrt{2}P_{l}] = i\mathbb I.
\eeq 

The eigenvectors of $B_{[{\bf k}]_{l}}$ are given, with 
$|\Phi^{[\bf k]}_{0_{l}}\rangle  = e^{i \sqrt{2} \frac{g_{[\bf k]}}{\omega_{l}}P_{l}}|0\rangle,$ by 
\beq {\label{eigvec}}
|\Phi^{[\bf k]}_{n_{l}}\rangle = e^{i \sqrt{2}\frac{g_{[\bf k]}}{\omega_{l}}P_{l}}|\phi_{n_{l}}\rangle 
= \frac{(A^{\dag}_{[{\bf k}]_{l}})^{n_{l}}}{\sqrt{n_{l} !}f_{l}(n_{l}) !}|\Phi^{[\bf k]}_{0_{l}}\rangle, \quad 
A_{[{\bf k}]_{l}} = e^{i \sqrt{2} {\frac{g_{[\bf k]}}{\omega_{l}}}P_{l}}a_{l}f_{l}(N_{l})e^{-i\sqrt{2}\frac{g_{[\bf k]}}{\omega_{l}}P_{l}} 
\eeq 
while for  the operators   $B_{[\bf k]}$  the eigenvalues  $E^{[\bf k]}_{\bf n}$ and eigenfunctions  $|\Phi^{[\bf k]}_{\bf n}\rangle $ are 
computed as follows: 
\beq
E^{[\bf k]}_{\bf n} = \sum_{l=1}^{N_B}\omega_{l}n_{l}f_{l}^{2}(n_{l}) + \epsilon_{[\bf k]} - g^{2}_{[\bf k]}\sum_{l=1}^{N_B}\frac{1}{\omega_{l}}, 
\quad \quad |\Phi^{[\bf k]}_{\bf n}\rangle = \bigotimes_{l=1}^{N_B}  |\Phi^{[\bf k]}_{n_l}\rangle.
\eeq

Let us denote the {\it{duals}} of the operators  $B_{[{\bf k}]_{l}}, A_{[{\bf k}]_{l}}, A^{\dag}_{[{\bf k}]_{l}}$ 
by $B'_{[{\bf k}]_{l}}, A'_{[{\bf k}]_{l}}, A'^{\dag}_{[{\bf k}]_{l}},$ respectively, such that 
\beq
B'_{[{\bf k}]_{l}} := \omega_{l}A'^{\dag}_{[{\bf k}]_{l}}A'_{[{\bf k}]_{l}} + \frac{\epsilon_{[\bf k]}}{N_{B}} 
- \frac{g^{2}_{[\bf k]}}{\omega_{l}},  \qquad A'_{[{\bf k}]_{l}} := a_{l}f_{l}(N_{l})^{-1} + \frac{g_{[\bf k]}}{\omega_{l}},
\eeq 
and
define, on the separable Hilbert space  ${\mathfrak H}_{l},$ 
%by $P'_{l}$, 
the {\it{dual}} of the momentum operator  $P_{l}$ by 
\beq 
P'_{l} := \frac{a_{l}f_{l}(N_{l})^{-1} - f_{l}(N_{l})a^{\dag}_{l}}{i \sqrt{2}} \quad \mbox{with} \quad 
[a_{l}f_{l}(N_{l})^{-1},\sqrt{2}P'_{l}] = i\mathbb I. 
\eeq

Then, the eigenvectors of  $B'_{[{\bf k}]_{l}}$ are given,  with 
$|\Phi'^{[\bf k]}_{0_{l}}\rangle  = e^{i \sqrt{2} \frac{g_{[\bf k]}}{\omega_{l}}P'_{l}}|0\rangle,$ by 
\beq{\label{eigvec00}} 
|\Phi'^{[\bf k]}_{n_{l}}\rangle = e^{i \sqrt{2}\frac{g_{[\bf k]}}{\omega_{l}}P'_{l}}|\phi_{n_{l}}\rangle 
= \frac{(A'^{\dag}_{[{\bf k}]_{l}})^{n_{l}}}{\sqrt{n_{l}} !} f_{l}(n_{l}) !|\Phi'^{[\bf k]}_{0_{l}}\rangle, \quad 
A'_{[{\bf k}]_{l}} = e^{i \sqrt{2} {\frac{g_{[\bf k]}}{\omega_{l}}}P'_{l}}a_{l}f_{l}(N_{l})^{-1}
e^{-i\sqrt{2}\frac{g_{[\bf k]}}{\omega_{l}}P'_{l}}\nonumber \\
\eeq
%Then to  the operators  $B'_{[\bf k]}$ 
corresponding to the eigenvalues 
\beq
E'^{[\bf k]}_{\bf n} = \sum_{l=1}^{N_B}\omega_{l}n_{l}f_{l}^{-2}(n_{l}) + \epsilon_{[\bf k]} 
- g^{2}_{[\bf k]}\sum_{l=1}^{N_B}\frac{1}{\omega_{l}}.
\eeq

It then comes that  the eigenvectors of $H$ associated to the eigenvalues $E^{[\bf k]}_{\bf n} $ are given by
\beq
|\varphi^{[{\bf k}]}_{\bf n}\rangle = |\Phi^{[\bf k]}_{\bf n}\rangle \otimes |\Psi_{[{\bf k}]}\rangle, \qquad
|\Psi_{[{\bf k}]}\rangle = (c^{\dag}_1)^{k_1}(c^{\dag}_2)^{k_2}\cdots
(c^{\dag}_j)^{k_j} \cdots (c^{\dag}_{M})^{k_M} |\Psi_{(0,0,\dots, 0)}\rangle.
\eeq

Since the operators $e^{i \sqrt{2}\frac{g_{[\bf k]}}{\omega_{l}}P_{l}}$,  
$e^{i \sqrt{2}\frac{g_{[\bf k]}}{\omega_{l}}P'_{l}}$ are unitary and act on $\mathfrak H_{l}$, 
the sets $\{|\Phi^{[\bf k ]}_{n_{l}}\rangle\}^{\infty}_{n_{l}=0}$ 
and $\{|\Phi'^{[\bf k ]}_{n_{l}}\rangle\}^{\infty}_{n_{l}=0}$ are both basis vectors of $\mathfrak H_{l}$.

\subsection{Construction of the nonlinear coherent states (NCS)}
From  the relations  (\ref{eigvec}) and (\ref{eigvec00}), we arrive at the following definitions:
\beq\label{fact00}
 \{n_{l}\} !:=n_{l} ![f_{l}(n_{l}) !]^{2}, \qquad \left\{n_{l} \right\}_{d}!:=\frac{n_{l} !}{[f_{l}(n_{l}) !]^{2}}.
\eeq
For all $l$,   we consider now,  on ${\mathfrak H}_{l},$ the displacement operators, related to the dual pair
 ${\mathfrak H}^{F}_{l}, {\mathfrak H}^{F^{-1}}_{l}$, as: 
% $D_{l}(z_{l})$ and $D'_{l}(z_{l})$ 
\beq{\label{dom1}} 
D_{l}(z_{l}) := e^{\imath \sqrt{2}\frac{g_{[\bf k]}}{\omega_{l}}P_{l}} e^{z_{l}A'^{\dag}_{l}-\bar{z}_{l}A_{l}}, \, z_{l} 
\in \mathcal D_{l}=\{z_{l}\in \C/ |z_{l}|<L=1/{\sqrt{\rho}}\},
\eeq
\beq{\label{dom2}}
D'_{l}(z_{l}) := e^{\imath \sqrt{2} \frac{g_{[\bf k]}}{\omega_{l}}P'_{l}}e^{z_{l}A^{\dag}_{l}-\bar{z}_{l}A'_{l}},\,z_{l} 
\in \tilde{\mathcal D_{l}}=\{z_{l}\in \C/ |z_{l}|<\tilde{L}=1/{\sqrt{\tilde{\rho}}}\}.
\eeq
Since the operators  $e^{\imath \sqrt{2}\frac{g_{[\bf k]}}{\omega_{l}}P_{l}}$ and $e^{\imath \sqrt{2} \frac{g_{[\bf k]}}{\omega_{l}}P'_{l}}$ 
are unitary on $\mathfrak H_{l}$, the above operators   are also unitary on $\mathfrak H_{l},$ 
and, in view of the relations 
\beq
D_{l}(z_{l})D_{l}(z'_{l}) &=& e^{\imath 2\sqrt{2}\frac{g_{[\bf k]}}{\omega_{l}}P_{l}} e^{\imath Im({\bar{z_{l}}z'_{l})}}D_{l}(z_{l} + z'_{l}), \cr
\cr
D'_{l}(z_{l})D'_{l}(z'_{l}) &=& e^{\imath 2\sqrt{2}\frac{g_{[\bf k]}}{\omega_{l}}P'_{l}} e^{\imath Im({\bar{z_{l}}z'_{l})}}D'_{l}(z_{l} + z'_{l})
\eeq
which hold if $z_{l},z'_{l},z_{l}+z'_{l} \in \mathcal D_{l}$ (resp. $\in \tilde{\mathcal D_{l}} $), 
they realize a unitary projective representation of the Weyl-Heisenberg group on $\mathfrak H_{l}$.

By using the  Baker-Campbell-Hausdorff identity, we obtain  
\beq{\label{ncs00}} 
D_{l}(z_{l}) = e^{\imath \sqrt{2}\frac{g_{[\bf k]}}{\omega_{l}}P_{l}} e^{-\frac{1}{2}|z_{l}|^{2}}e^{z_{l}A'^{\dag}_{l}}e^{-\bar{z}_{l}A_{l}},
\eeq
\beq{\label{ncs01}}
D'_{l}(z_{l}) = e^{\imath \sqrt{2} \frac{g_{[\bf k]}}{\omega_{l}}P'_{l}} e^{-\frac{1}{2}|z_{l}|^{2}}e^{z_{l}A^{\dag}_{l}}e^{-\bar{z}_{l}A'_{l}} .
\eeq
From (\ref{ncs00}),  for all $z_{l} \in \mathcal D_{l},$ the NCS
\beq{\label{coh1}}
D_{l}(z_{l})|0\rangle_{l} 
%&=& e^{\imath \sqrt{2}\frac{g_{[\bf k]}}{\omega_{l}}P_{l}} e^{-\frac{1}{2}|z_{l}|^{2}}e^{z_{l}A'^{\dag}_{l}}
%e^{-\bar{z}_{l}A_{l}}|0\rangle_{l} \cr 
                          = e^{-\frac{1}{2}|z_{l}|^{2}}
e^{\imath \sqrt{2}\frac{g_{[\bf k]}}{\omega_{l}}P_{l}}\sum_{n_{l}=0}^{\infty}\frac{z^{n_{l}}_{l}}
{\sqrt{\left\{n_{l} \right\}!} }|\phi_{n_{l}}\rangle 
\eeq
 become in the domain $\mathcal D_{l}$ given in (\ref{dom1}):
\beq{\label{coh2}}
D_{l}(z_{l})|0\rangle_{l} := \left(\mathcal N(|z_{l}|^{2})\right)^{-1/2}
\sum_{n_{l}=0}^{\infty}\frac{z^{n_{l}}_{l}}{\sqrt{\left\{n_{l} \right\}!} }|\Phi^{[\bf k ]}_{n_{l}}\rangle 
\eeq
with the normalization factor, by using (\ref{fact00}),   given by 
\beq
\mathcal N(|z_{l}|^{2})   = 
\sum_{n_{l}=0}^{\infty}\frac{(\bar{z}_{l}z_{l})^{n_{l}}}{\left\{n_{l} \right\}!}.
\eeq
These vectors   correspond to the well-known
 {\it NCS} of quantum optics \cite{man'ko},  related to the basis vectors 
$\{|\Phi^{[\bf k ]}_{n_{l}}\rangle\}^{\infty}_{n_{l}=0}$ of the separable Hilbert space $\mathfrak H_{l}$. 

Similarly, from  (\ref{ncs01}), we determine, for all $z_{l} \in \tilde{\mathcal D_{l}},$ the NCS
 
\beq{\label{coh4}} 
D'_{l}(z_{l})|0\rangle_{l} &=& \left(\mathcal N'(|z_{l}|^{2})\right)^{-1/2}
\sum_{n_{l}=0}^{\infty}\frac{z^{n_{l}}_{l}  }
{\sqrt{\left\{n_{l} \right\}_d!}}|\Phi'^{[\bf k]}_{n_{l}}\rangle
\eeq
with
\beq
\mathcal N'(|z_{l}|^{2})  = 
\sum_{n_{l}=0}^{\infty}\frac{(\bar{z}_{l}z_{l})^{n_{l}}}{\left\{n_{l} \right\}_{d}!}.
\eeq
% These vectors correspond to the so-called NCS 
%constructed by Roy {\it et  al} \cite{roy-roy}. 

Setting now 
\beq
|0\rangle = |0\rangle_{1} \otimes |0\rangle_{2} \otimes \cdots \otimes |0\rangle_{l} \cdots \otimes |0\rangle_{N_{B}} 
\eeq
and considering   the displacement operators $D_{\bf n}({\bf z})$ and $D'_{\bf n}({\bf z}):$
\beq
 D_{\bf n}({\bf z}):=\bigotimes_{l=1}^{N_{{B}}}D_{l}(z_{l}),\qquad 
D'_{\bf n}({\bf z}):=\bigotimes_{l=1}^{N_{{B}}}D'_{l}(z_{l}),
\eeq
 the action of $ D_{\bf n}({\bf z})$ on $|0\rangle$ can be realized as:
\beq
 D_{\bf n}({\bf z})|0\rangle &=& \bigotimes_{l=1}^{N_{{B}}}D_{l}(z_{l})(|0\rangle_{1} \otimes 
|0\rangle_{2} 
\otimes 
\cdots \otimes |0\rangle_{l} \otimes \cdots \otimes |0\rangle_{N_{B}})\cr
                         &=& D_{1}(z_{1})|0\rangle_{1} \otimes D_{2}(z_{2})|0\rangle_{2} \otimes \cdots 
\otimes D_{l}(z_{l})|0\rangle_{l} \otimes \cdots \otimes  D_{N_{B}}(z_{N_{B}})|0\rangle_{N_{B}} \cr
                         &=& \left(\mathcal N(|z_{1}|^{2})\right)^{-1/2}
\sum_{n_{1}=0}^{\infty}\frac{z^{n_{1}}_{1}}{\sqrt{\left\{n_{1} \right\}!} }\, \left(\mathcal N(|z_{2}|^{2})\right)^{-1/2}
\sum_{n_{2}=0}^{\infty}\frac{z^{n_{2}}_{2}}{\sqrt{\left\{n_{2} \right\}!} } \cr
 &&\cdots \left(\mathcal N(|z_{l}|^{2})\right)^{-1/2}
\sum_{n_{l}=0}^{\infty}\frac{z^{n_{l}}_{l}}{\sqrt{\left\{n_{l} \right\}!} } 
\cdots \left(\mathcal N(|z_{N_{B}}|^{2})\right)^{-1/2} \sum_{n_{N_{B}}=0}^{\infty}\frac{z^{n_{N_{B}}}_{N_{B}}}{\sqrt{\left\{n_{N_{B}} \right\}!} }\cr
&&\times
(|\Phi^{[\bf k]}_{n_{1}}\rangle \otimes |\Phi^{[\bf k]}_{n_{2}}\rangle \otimes \cdots\otimes  |\Phi^{[\bf k]}_{n_{l}}\rangle \otimes \cdots 
\otimes |\Phi^{[\bf k]}_{n_{N_{B}}}\rangle ). \nonumber
\\
\eeq
Putting
\beq{\label{write}}
\mathcal N(|{\bf z}|^{2}) := \prod_{l=1}^{N_{B}}\mathcal N_{l}(|z_{l}|^{2}), \qquad
{\bf z}^{\bf n}:= \prod_{l=1}^{N_{B}} z^{n_{l}}_{l},
\eeq
and
\beq
{\bf n} !:= n_{1} !n_{2} ! \cdots n_{N_{B}} !, \qquad f({\bf n}) !:= f(n_{1}) !f(n_{2}) !\cdots f(n_{N_{B}}) !,
\eeq
 we derive the NCS as follows:
\beq\label{ncsvs00}
|\eta_{\bf z} \rangle := D_{\bf n}({\bf z})|0\rangle =\left(\mathcal N(|{\bf z}|^{2})\right)^{-1/2}
\sum_{{\bf n}=0}^{\infty}\frac{{\bf z}^{\bf n}}{\sqrt{\left\{{\bf n}\right\} !} }
|\Phi^{[\bf k]}_{\bf n}\rangle, \qquad    {\bf z} \in \mathcal D
\eeq
and  
\beq\label{ncsvs01}
|\eta'_{\bf z} \rangle := D'_{\bf n}({\bf z})|0\rangle =\left(\mathcal N'(|{\bf z}|^{2})\right)^{-1/2}
\sum_{{\bf n}=0}^{\infty}\frac{{\bf z}^{\bf n} }{\sqrt{\left\{{\bf n}\right\}_d !}}
|\Phi'^{[\bf k]}_{\bf n}\rangle, \qquad    {\bf z} \in \tilde{\mathcal D}.
\eeq
\begin{rmk}  
In the previous definitions, for each $l, l = 1,2,\dots,N_{B}$,  the variables $z_{l}$   
are assumed to be mutually independent.
\end{rmk}

On $ \mathfrak H$ spanned by the bosonic eigenvectors $|\Phi^{[\bf k]}_{\bf n}\rangle$ and $|\Phi'^{[\bf k]}_{\bf n}\rangle$, 
we have  the following resolutions of the identity:
\beq\label{relat02}
\int_{\mathcal D} |\eta_{\bf z} \rangle \langle \eta_{\bf z} | \,  
d\mu ({\bf z},{\bar{\bf z}})  \mathcal N(|{\bf z}|^{2})  
=   I_{\mathfrak H}, \\
\int_{\tilde{\mathcal D}}
|\eta'_{\bf z} \rangle \langle \eta'_{\bf z} |  \,
d\mu ({\bf z},{\bar{\bf z}})   \mathcal N'(|{\bf z}|^{2})  
= I_{\mathfrak H}.
\eeq 
The measure $ d\mu({\bf z},{\bar{\bf z}})$ is such that $d\mu({\bf z},{\bar{\bf z}})= 
 d\lambda(r)d\theta, \, {\bf z}=re^{\imath \theta}$,  where 
$d\lambda(r)$ is determined through the moment problem:
\beq
2\pi \int_{0}^{L} r^{2{\bf n}} d\lambda(r) = \left\{{\bf n}\right\} !, \qquad  {\bf n} = 0, 1, 2, \cdots
\eeq
in the case of the vectors $D_{\bf n}({\bf z})|0\rangle;$ for the vectors $D'_{\bf n}({\bf z})|0\rangle$, with 
$d\mu({\bf z},{\bar{\bf z}})=d\varrho(r)d\theta$,  
\beq
2\pi \int_{0}^{\tilde L} r^{2{\bf n}} d\varrho(r) = \left\{{\bf n}\right\}_{d} !, \qquad  {\bf n} = 0, 1, 2, \cdots.
\eeq

\section{ Probability density computation}
This section discusses the probability density and its time evolution  evaluated in the constructed NCS basis.

We start computing the quantity

\beq\label{kern00}
\langle \eta_{\bf z}|\eta_{\bf z_0} \rangle &=&  \left[\mathcal N(|{\bf z}|^{2})\mathcal N(|{\bf z_0}|^{2})\right]^{-1/2}
\sum_{{\bf n}=0}^{\infty}
\frac{{(\bar{\bf z}} {\bf z_0})^{\bf n} }{\left\{{\bf n}\right\}  !} \cr
&=& \frac{\mathcal N(\bar{\bf z}{\bf z_0})}{\sqrt{\mathcal N(|{\bf z}|^{2})\mathcal N(|{\bf z_0}|^{2})}}
\eeq

giving  

\beq
|\langle \eta_{\bf z}|\eta_{\bf z_0} \rangle|^2 =
\frac{|\mathcal N(\bar{\bf z}{\bf z_0})|^2}{\mathcal N(|{\bf z}|^{2}) \mathcal N(|{\bf z_0}|^{2})}, \qquad |\mathcal N(\bar{\bf z}{\bf z_0})|^2 := 
\mathcal N(\bar{\bf z}{\bf z_0})\mathcal N({\bf z}\bar{\bf z}_{0}).
\eeq

Then, the  probability density is defined as the map

\beq\label{dens00}
\mathcal D &\rightarrow  \R_+, \quad\quad
%\cr
{\bf z} \mapsto& \varrho_{{\bf z_0}}({\bf z}) := |\langle \eta_{\bf z}|\eta_{\bf z_0} \rangle|^2  = 
\frac{|\mathcal N(\bar{\bf z}{\bf z_0})|^2}{\mathcal N(|{\bf z}|^{2}) \mathcal N(|{\bf z_0}|^{2})}.
\eeq

Consider the following spectral decomposition in the eigenstates $|\Phi^{[{\bf{k}}]}_{{\bf{n}}}\rangle$ given by 
\beq
\tilde H = \sum_{ {\bf{n}} =0}^{\infty}\Omega_{N_B} \mathcal E_{{\bf{n}}} \, 
|\Phi^{[{\bf{k}}]}_{{\bf{n}}}\rangle \langle \Phi^{[{\bf{k}}]}_{{\bf{n}}}|,\quad \mathcal E_{{\bf{n}}} = \sum_{l=1}^{N_B}n_{l}f_{l}(n_{l}) 
\eeq
with 
\beq{\label{omega}} \Omega_{N_B} :=  \left(\sum_{l=1}^{N_B}\omega_{l}
n_{l}f_{l}(n_{l})\right)\left/ \left(\sum_{l=1}^{N_B} n_{l}f_{l}(n_{l})\right)\right.. \eeq

Let
\beq 
{\bf{J}}^{{\bf{n}}/2} := \prod_{l=1}^{N_B} J^{n_l/2}_{l}, \quad \varepsilon_{\bf n}:=\left(
                              n_{1}f_1(n_{1}), n_{2}f_2(n_{2}),  \cdots,
n_{N_B}f_{N_B}(n_{N_B})
\right), \quad \gamma :=
                              \left(
                              \begin{array}{c}
                              \gamma_{1}\\
                               \gamma_{2} \\
                               \vdots \\
                                \gamma_{N_B}\\
                               \end{array}
                                \right), 
\eeq
where ${\bf z_0}^{\bf n}  := {\bf{J}_0}^{{\bf{n}}/2}e^{-\imath \varepsilon_{\bf n}\gamma_0}$, such that we get 
  
\beq
e^{-\imath \tilde H t}|\eta_{\bf z_0} \rangle 
= 
\left(\mathcal N(|{\bf z_0}|^{2})\right)^{-1/2} \sum_{{\bf n}=0}^{\infty}
\frac{ {\bf{J}_0}^{{\bf{n}}/2}e^{-\imath \varepsilon_{\bf n}\gamma_0}}{\sqrt{\left\{{\bf n}\right\}!}}
  e^{-\imath \left(\Omega_{N_B}\mathcal E_{{\bf{n}}} \right)t}|\Phi^{[{\bf{k}}]}_{{\bf{n}}}\rangle.
\eeq
Since 
\beq{\label{mat00}}
\varepsilon_{{\bf n}} \beta = \left(
                              n_{1}f_1(n_{1}), n_{2}f_2(n_{2}),  \cdots,
n_{N_B}f_{N_B}(n_{N_B})
\right)\left(
                      \begin{array}{c}
                         1 \\
                        \vdots \\
                         1 \\
                        \vdots \\
                        1 \\
                       \end{array}
                        \right)  = \sum_{l=1}^{N_B}n_{l}f_{l}(n_{l}) = \mathcal E_{{\bf n}},
\eeq
we have
\beq
e^{-\imath \tilde H t}|\eta_{\bf z_0} \rangle 
=
|{\bf J}_0,\gamma_0 +
\Omega_{N_B}t\beta\rangle =  |\eta_{\bf z_0(t)} \rangle, \qquad  ({\bf z_0}(t))^{\bf{n}} := 
{\bf{J}_0}^{{\bf{n}}/2}e^{-\imath \varepsilon_{\bf n}(\gamma_0+\Omega_{N_B}t\beta)}
\eeq 

and thereby 

\beq
|\langle \eta_{\bf z}|e^{- i  \tilde H t}|\eta_{\bf z_0} \rangle |^2 =
\frac{|\mathcal N(\bar{\bf z}{\bf z_0}(t))|^2}{\mathcal N(|{\bf z}|^{2})\mathcal N(|{\bf z_0}|^{2})}.
\eeq

Thus, the  time evolution behavior of $\varrho_{z_0}(z)$ is provided by the mapping

\beq\label{dens01}
{\bf z} \mapsto \varrho_{{\bf z_0}}({\bf z},t) := |\langle \eta_{\bf z}|e^{- i  \tilde H t}|\eta_{\bf z_0} \rangle |^2 
=\frac{|\mathcal N(\bar{\bf z}{\bf z_0}(t))|^2}{\mathcal N(|{\bf z}|^{2})\mathcal N(|{\bf z_0}|^{2})} 
\eeq

while  the dynamics of the  NCS $|\eta_{\bf z}\rangle$ is governed by the relation: 

\beq\label{dens02}
|\eta_{\bf z}; t\rangle  = e^{- i  \tilde H t}|\eta_{\bf z}\rangle = 
|\eta_{\bf z(t)} \rangle, \qquad  ({\bf z}(t))^{\bf{n}} := 
{\bf{J}}^{{\bf{n}}/2}e^{-\imath \varepsilon_{\bf n}(\gamma +\Omega_{N_B}t\beta)}.
\eeq

Finally, as

\beq
\langle \eta'_{\bf z}|\eta'_{\bf z_0} \rangle =  \left[\mathcal N(|{\bf z}|^{2})\mathcal N(|{\bf z_0}|^{2})\right]^{-1/2}
\sum_{{\bf n}=0}^{\infty}
\frac{{(\bar{\bf z}} {\bf z_0})^{\bf n}  }{\left\{{\bf n}\right\}_{d} !}, 
\eeq

the NCS $|\eta'_{\bf z}\rangle$ obey the same relations  as (\ref{dens01}) and (\ref{dens02}).

\section{Reproducing kernel}
In this section,  we discuss another important property of the NCS, i.e. the reproducing 
kernel on the Hilbert space $\mathfrak H$, based on
 the completeness relation (\ref{relat02}).

From the expression (\ref{kern00}), we introduce the  quantity 

\beq\label{kern01}
\mathcal K({\bf z}, {\bf z'}):=
 \langle \eta_{\bf z'}|\eta_{\bf z} \rangle  =  \frac{\mathcal N(\bar{\bf z'}{\bf z})}{\sqrt{\mathcal N(|{\bf z'}|^{2})\mathcal N(|{\bf z}|^{2})}}
\eeq

standing for a  reproducing kernel \cite{ali-antoine-gazeau}. Indeed,

\bpro\label{kernelprop} 

The following properties are satisfied by the function $\mathcal K$ on the Hilbert space $\mathfrak H$:

\bitem

\item [(i)] Hermiticity  
\beq
\mathcal K({\bf z}, {\bf z'}) = \overline{\mathcal K({\bf z'}, {\bf z})};
\eeq

\item [(ii)] Positivity
\beq
\mathcal K({\bf z}, {\bf z})  >  0;
\eeq

\item [(iii)]  Idempotence

\beq\label{kern02}
\int_{\mathcal D}  d\mu ({\bf z''},{\bar{\bf z''}}) \mathcal K({\bf z}, {\bf z''})\mathcal K({\bf z''}, {\bf z'}) = 
\mathcal K({\bf z}, {\bf z'}).
\eeq

\eitem

\epro

{\bf Proof.} See Appendix.

$\hfill{\square}$

%{\color{blue}

%From (\ref{resolv}),
For any $|\Psi\rangle \in \mathfrak H$, we have 

\beq
|\Psi \rangle  =   \int_{\mathcal D} d\mu ({\bf z},{\bar{\bf z}})  \Psi({\bf z})  |\eta_{\bf z}\rangle 
\eeq
where $\Psi({\bf z}) :=  \langle  \eta_{\bf z}|\Psi\rangle $. 
The following reproducing property 

\beq\label{relat03}
\Psi({\bf z}) =  \int_{\mathcal D} d\mu ({\bf z'},{\bar{\bf z'}}) \mathcal K({\bf z},{\bf z'}) \Psi({\bf z'})
\eeq
is also verified.
 
%}

\section{Diagonal representation of the density matrix}

In this section,  the statistical properties of the NCS are investigated. 
We use the results issued from the previous section   as   key ingredients  to construct a  Glauber-Sudarshan $P$-representation 
 of the density matrix following \cite{brif-aryeh, cahill-glauber,  gazbook09}, and also \cite{parthasarathy-sridhar}, where  a  $q$-analogue 
has been performed.

The    density matrix, for a fixed $[\bf k],$ is given, in terms of the bosonic states,  as
\beq\label{densmat}
\rho_{[{\bf k}]} =  \sum_{{\bf n}, {\bf m}=0}^{\infty} \rho_{[{\bf k}]}({\bf n}, {\bf m})
|\Phi^{[{\bf k}]}_{\bf n}\rangle \langle \Phi^{[{\bf k}]}_{\bf m}| 
\eeq
where the $\rho_{[{\bf k}]}({\bf n}, {\bf m})$ are the matrix elements.

In a matrix form, we get, by using   (\ref{densmat}),   the following expression for $\hat \rho:$

\beq
\hat \rho
&=&\sum_{[{\bf k}]\in \Gamma} \sum_{{\bf n}, {\bf m}=0}^{\infty}\rho_{[{\bf k}]}({\bf n}, {\bf m})
|\Psi_{[\bf k]}\otimes\Phi^{[\bf k]}_{[{\bf{n}}]}\rangle
\langle\Psi_{[\bf k]}\otimes\Phi^{[\bf k]}_{[{\bf{m}}]}| 
\eeq
which takes on the Hilbert space $\C^{2^{M}} \otimes \mathfrak H$ the form
\beq
\hat \rho
&=& diag(\rho_{(0,0, \dots, 0,0)}, \rho_{(1,0, \dots 0,0)}, 
\cdots, \rho_{[{\bf k}]}, \cdots, \rho_{(1,1, \dots, 1,1)})
\eeq

where   $[{\bf{n}}]: = n_1, n_2, \dots,  n_{N_B} $ and   
$\mathfrak H = \bigoplus_{[{\bf k}] \in \Gamma}\mathfrak H_{[{\bf k}]}$, $\mathfrak H_{[{\bf k}]}$ 
is the subspace of $\mathfrak H $ spanned by the vectors 
$|\Phi^{[\bf k]}_{[{\bf{n}}]}\rangle$ 
with  $\sum_{[{\bf k}]\in \Gamma}|\Psi_{[\bf k]}\rangle \langle \Psi_{[\bf k]}| = \mathbb I_{2^{M}}$ with  $\Gamma$ 
 defined in (\ref{set00}).

%}

Keeping in mind  the radial parametrization ${\bf z}=re^{\imath \theta}$ we get   the following quantity 
$\mathcal N(|{\bf z}|^{2})|re^{\imath \theta}\rangle \langle re^{\imath \theta}|$:

\beq
\mathcal N(|{\bf z}|^{2})|re^{\imath \theta}\rangle \langle re^{\imath \theta}|=
  \sum_{{\bf n}, {\bf m}=0}^{\infty} 
\frac{{r}^{{\bf m}+{\bf n}} e^{\imath ({\bf n}-{\bf m})\theta} }{\sqrt{\left\{{\bf n}\right\}!\left\{{\bf m}\right\}!}}|\Phi^{[\bf k]}_{\bf n}\rangle
\langle \Phi^{[\bf k]}_{\bf m}|
\eeq
which leads to the integral

\beq
\int_{0}^{2\pi}\frac{d\theta}{2\pi} 
e^{\imath {\bf l} \theta}\mathcal N(|{\bf z}|^{2})|re^{\imath \theta}\rangle \langle re^{\imath \theta}|&=& 
\sum_{{\bf n}, {\bf m}=0}^{\infty}\frac{{r}^{{\bf m}+{\bf n}}}{\sqrt{\left\{{\bf n}\right\}!
\left\{{\bf m}\right\}!}}\delta({\bf n}-{\bf m} +  {\bf l})  |\Phi^{[\bf k]}_{\bf n}\rangle
\langle \Phi^{[\bf k]}_{\bf m}|.
\eeq 

We obtain the expression  

\beq\label{exps00}
\left(\frac{d}{dr}\right)^{\bf q}
\int_{0}^{2\pi}\frac{d\theta}{2\pi}e^{\imath{\bf l}\theta}\mathcal N(|{\bf z}|^{2})|re^{\imath \theta}\rangle \langle re^{\imath \theta}|
&=&  
\sum_{{\bf n}, {\bf m}=0}^{\infty}\frac{{r}^{{\bf m}+{\bf n}- {\bf q}}}{\sqrt{\left\{{\bf n}\right\}!
\left\{{\bf m}\right\}!}}\frac{({\bf m}+{\bf n}) !}{({\bf m}+{\bf n}- {\bf q}) !}\cr
&& \times \delta({\bf n}-{\bf m} +  {\bf l}) |\Phi^{[\bf k]}_{\bf n}\rangle
\langle \Phi^{[\bf k]}_{\bf m}|
\eeq

such that the following quantity

\beq
\left\{\left(\frac{d}{dr}\right)^{\bf q}
\int_{0}^{2\pi}\frac{d\theta}{2\pi}e^{\imath {\bf l}\theta}
\mathcal N(|{\bf z}|^{2})|re^{\imath \theta}\rangle \langle re^{\imath \theta}|\right\}_{r=0},
\eeq

where, at $r = 0$  in the
right-hand side of (\ref{exps00}), the term with $ {\bf m}+{\bf n}- {\bf q} =0$  alone survives, 
provides

\beq
|\Phi^{[\bf k]}_{\bf n}\rangle
\langle \Phi^{[\bf k]}_{\bf m}| =  \frac{\sqrt{\left\{{\bf n}\right\}!
\left\{
{\bf m}\right\}!}}{({\bf m}+{\bf n}) !}
%\left\{
\left(\frac{d}{dr}\right)^{{\bf m}+{\bf n}}
\left.\int_{0}^{2\pi}\frac{d\theta}{2\pi}e^{\imath ({\bf m}-{\bf n}) \theta}
\mathcal N(|{\bf z}|^{2})|re^{\imath \theta}\rangle \langle re^{\imath \theta}|\right\rvert_{r=0}.
\nonumber\\
\eeq

Then, a $f$-deformed   Glauber-Sudarshan $P$-representation 
 of the density matrix in the bosonic states basis is given by 

\beq
\rho_{[{\bf k}]} = \sum_{{\bf n}, {\bf m} =0}^{\infty} \rho_{[{\bf k}]}({\bf n}, {\bf m})  
\frac{\sqrt{\left\{{\bf n}\right\}!
\left\{{\bf m}\right\}!}}{({\bf m}+{\bf n}) !}
%\cr
%&& \times 
\left(\frac{d}{dr}\right)^{{\bf m}+{\bf n}}
\left.\int_{0}^{2\pi}\frac{d\theta}{2\pi}e^{\imath ({\bf m}-{\bf n}) \theta}
\mathcal N(|{\bf z}|^{2})|re^{\imath \theta}\rangle \langle re^{\imath \theta}|\right\rvert_{r=0}.
\nonumber\\
\eeq

In the case of the dual NCS $|\eta'_{\bf z}\rangle$, we get
%\beq
% \left(\frac{d}{dr}\right)^{{\bf m}+{\bf n}}
%\left.\int_{0}^{2\pi}\frac{d\theta}{2\pi}e^{\imath ({\bf m}-{\bf n}) \theta}
%\mathcal N(|{\bf z}|^{2})|re^{\imath \theta}\rangle \langle re^{\imath \theta}|\right\rvert_{r=0} =
%\frac{({\bf m}+{\bf n}) !}{\sqrt{\left\{{\bf n}\right\}_d!
%\left\{{\bf m}\right\}_d!}} |\Phi^{[\bf k]}_{\bf n}\rangle
%\langle \Phi^{[\bf k]}_{\bf m}| \nonumber \\
%\eeq

%with 

\beq
|\Phi^{[\bf k]}_{\bf n}\rangle
\langle \Phi^{[\bf k]}_{\bf m}| = \frac{\sqrt{\left\{{\bf n}\right\}_d!
\left\{{\bf m}\right\}_d!}}{({\bf m}+{\bf n}) !}\left(\frac{d}{dr}\right)^{{\bf m}+{\bf n}}
\left.\int_{0}^{2\pi}\frac{d\theta}{2\pi}e^{\imath ({\bf m}-{\bf n}) \theta}\mathcal N(|{\bf z}|^{2})|re^{\imath \theta}\rangle 
\langle re^{\imath \theta}|\right\rvert_{r=0},
%\nonumber 
%\\
\eeq

%\item  

and

\beq
\tilde\rho_{[{\bf k}]} = \sum_{{\bf n}, {\bf m} =0}^{\infty}  \tilde\rho_{[{\bf k}]}({\bf n}, {\bf m})  
\frac{\sqrt{\left\{{\bf n}\right\}_d!
\left\{{\bf m}\right\}_d!}}{({\bf m}+{\bf n}) !}\left(\frac{d}{dr}\right)^{{\bf m}+{\bf n}}
\left.\int_{0}^{2\pi}\frac{d\theta}{2\pi}e^{\imath ({\bf m}-{\bf n}) \theta}\mathcal N(|{\bf z}|^{2})|re^{\imath \theta}\rangle 
\langle re^{\imath \theta}|\right\rvert_{r=0}.\nonumber \\
\eeq

Using the   definition   (\ref{densmat}), the matrix elements of $\rho_{[{\bf k}]}$  computed with  two  NCS is found to be 

\beq\label{densmat05}
\langle \eta_{\bf z'} |\rho_{[{\bf k}]}|\eta_{\bf z} \rangle = \left[\mathcal N(|{\bf z'}|^{2}) \mathcal N(|{\bf z}|^{2})\right]^{-1/2}
\sum_{{\bf m}, {\bf n} =0}^{\infty} \rho_{[{\bf k}]}({\bf n}, {\bf m})
\frac{\bar{\bf z'}^{\bf n}}{\sqrt{\left\{{\bf n}\right\}  !}}
\frac{{\bf z}^{\bf m}}{\sqrt{\left\{{\bf m}\right\}  !}} 
\eeq

which becomes at ${\bf z'}  = {\bf z} $, 

%{\color{blue}

\beq
\langle \eta_{\bf z} |\rho_{[{\bf k}]}|\eta_{\bf z} \rangle = \left(\mathcal N(|{\bf z}|^{2})\right)^{-1}
\sum_{{\bf m}, {\bf n} =0}^{\infty} \rho_{[{\bf k}]}({\bf n}, {\bf m})
\frac{\bar{\bf z}^{\bf n}}{\sqrt{\left\{{\bf n}\right\}  !}}
\frac{{\bf z}^{\bf m}}{\sqrt{\left\{{\bf m}\right\}  !}}. 
\eeq

%}

The quantity $\langle \eta_{\bf z} |\rho_{[{\bf k}]}|\eta_{\bf z} \rangle$ 
is analog to the \textquotedblleft semi-classical \textquotedblright phase space 
distribution 
function  $\mu(x,p) =  \langle z|\rho|z \rangle$ 
 associated to the density matrix $\rho$ (here $\rho_{[{\bf k}]}$)  of the system which is normalized as $\int(dx dp/2\pi \hbar)\mu(x,p) = 1$, 
and often referred to 
as the Husimi distribution \cite{husimi}.

In terms of the NCS,  the density matrix is given by 

\beq\label{densmat01}
\rho_{[{\bf k}]} = \int_{\mathcal D} d\mu ({\bf z},{\bar{\bf z}})  \mathcal N(|{\bf z}|^{2})  
  P(|{\bf z}|^{2}) \, |\eta_{\bf z} \rangle \langle \eta_{\bf z} |, 
\eeq

where $P(|{\bf z}|^{2})$ satisfies the normalization condition

\beq\label{proba}
\int_{\mathcal D} d\mu ({\bf z},{\bar{\bf z}}) P(|{\bf z}|^{2}) = 1.
\eeq

Then,  

\beq
\langle \Phi^{[\bf k]}_{\bf n}|\rho_{[{\bf k}]}|\Phi^{[\bf k]}_{\bf n}\rangle 
%&=&  
%\int_{\mathcal D} d\mu ({\bf z},{\bar{\bf z}}) {  \mathcal N(|{\bf z}|^{2})}
% P({\bf z}) \, \langle \Phi^{[\bf k]}_{\bf n}|\eta_{\bf z} \rangle \langle \eta_{\bf z} |
%\Phi^{[\bf k]}_{\bf n}\rangle \cr
&=& \int_{\mathcal D} d\mu ({\bf z},{\bar{\bf z}})   \mathcal N(|{\bf z}|^{2}) 
 P(|{\bf z}|^{2}) \, |\langle \Phi^{[\bf k]}_{\bf n}|\eta_{\bf z} \rangle|^2
\eeq

where the quantity

\beq
|\langle \Phi^{[\bf k]}_{\bf n}|\eta_{\bf z} \rangle|^2 =\left(\mathcal N(|{\bf z}|^{2})\right)^{-1}
\frac{|{\bf z}|^{2\bf n}}{\left\{{\bf n}\right\}  !}
\eeq

corresponds to a $f$-deformed    photon number  equal to 
\beq
\mathcal P({\bf n}, {\bf z}) = e^{-|{\bf z}|^2} 
\frac{|{\bf z}|^{2\bf n}}{{\bf n}  !}, 
\eeq
when $f({\bf n}) \rightarrow 1$, which is a Poisson distribution. Therefore, we say that the NCS obey a 
$f$-Poisson distribution.

%Then, we get 

In the  VCS formalism, we write that

%{\color{blue}

\beq\label{matdensmat}
\hat \rho &=&\sum_{[{\bf k}]\in \Gamma}\int_{\mathcal  D^{2^{M}}} d\mu(\mathfrak Z, \bar{\mathfrak Z}) 
  \mathcal N(|{\bf z_{[{\bf k}]}}|^{2})   P(|{\bf z_{[{\bf k}]}}|^2)
|\eta_{\mathfrak Z};[{\bf k}]\rangle 
\langle \eta_{\mathfrak  Z};[{\bf k}]|
\eeq
where

\beq
|\eta_{\mathfrak Z};[{\bf k}]\rangle  = \left(
                      \begin{array}{c}
                         0 \\
                        \vdots \\
                         0 \\
                         |\eta_{{\bf z}_{[{\bf k}]}}\rangle  \\
                        0 \\
                        \vdots \\
                        0 \\
                       \end{array}
                        \right)
\eeq

coincide,  in the special case of
 $f(N_i) = 1 \, (1 \leq i \leq N_B)$,  with  the VCS developed in \cite{a-hk}, and     

\beq
|\eta_{{\bf z}_{[{\bf k}]}} \rangle =: \left(\mathcal N(|{\bf z_{[{\bf k}]}}|^{2})\right)^{-1/2}
\sum_{{\bf n}=0}^{\infty}\frac{{\bf z}^{\bf n}_{[{\bf k}]}}{\sqrt{\left\{{\bf n}\right\} !} }
|\Psi_{[\bf k]}\otimes\Phi^{[\bf k]}_{[{\bf{n}}]}\rangle
\eeq

such that,  with respect to the measure $d\mu(\mathfrak Z, \bar{\mathfrak Z})  
=     \prod_{[\bf k] \in \Gamma} d\mu({\bf z_{[{\bf k}]}}, {\bar{\bf z}_{[{\bf k}]}})$,  
the resolution of the identity   provided as in \cite{a-hk},

\beq
\sum_{[{\bf k}]\in \Gamma}\int_{\mathcal  D^{2^{M}}} d\mu(\mathfrak Z, \bar{\mathfrak Z}) 
  \mathcal N(|{\bf z_{[{\bf k}]}}|^{2}) 
|\eta_{\mathfrak Z};[{\bf k}]\rangle 
\langle \eta_{\mathfrak  Z};[{\bf k}]|
%\crcr
&=& \sum_{[{\bf k}]\in \Gamma}\sum_{{[\bf n]}=0}^{\infty}|\Psi_{[\bf k]}\otimes\Phi^{[\bf k]}_{[{\bf{n}}]}\rangle  
\langle\Psi_{[\bf k]}\otimes\Phi^{[\bf k]}_{[{\bf{n}}]}| \cr
&=& \mathbb I_{2^{M}} \otimes I_{\mathfrak H},
\eeq
is satisfied.

%}

%{\color{blue}
Keeping in mind the following quantity

\beq
|\eta_{\bf z}\rangle \langle \eta_{\bf z}| = \left(\mathcal N(|{\bf z}|^{2})\right)^{-1} 
 \sum_{{\bf j}, {\bf p}=0}^{\infty} \frac{{\bf z}^{\bf j}}{\sqrt{\left\{{\bf j}\right\}!} }
\frac{\bar{\bf z}^{\bf p}}{\sqrt{\left\{{\bf p}\right\}!} }|\Phi^{[\bf k]}_{\bf j}\rangle \langle \Phi^{[\bf k]}_{\bf p}| 
\eeq

and using the relation (\ref{densmat01}), the density matrix can be explicitly computed as 

\beq\label{densmat02}
\rho_{[{\bf k}]} =   \sum_{{\bf j}, {\bf p}=0}^{\infty} \int_{\mathcal D}  d\mu ({\bf z},{\bar{\bf z}}) P(|{\bf z}|^{2}) \,
\left\{ 
\frac{{\bf z}^{\bf j}}{\sqrt{\left\{{\bf j}\right\}!} }
\frac{\bar{\bf z}^{\bf p}}{\sqrt{\left\{{\bf p}\right\}!} }|\Phi^{[\bf k]}_{\bf j}\rangle \langle \Phi^{[\bf k]}_{\bf p}|\right\}.
\eeq

From (\ref{densmat}) and (\ref{densmat02}) together, it follows that
 
\beq\label{densmat03}
\langle \Phi^{[\bf k]}_{\bf n}|\rho_{[{\bf k}]}|\Phi^{[\bf k]}_{\bf n}\rangle := \rho_{[{\bf k}]}({\bf n}, {\bf n}) = 
\int_{\mathcal D} d\mu ({\bf z},{\bar{\bf z}})  P(|{\bf z}|^{2}) \,\
\frac{|{\bf z}|^{2\bf n}}{\left\{{\bf n}\right\}  !},
\eeq
 and the matrix elements of the density operator evaluated in  the bosonic 
states $|\Phi^{[\bf k]}_{\bf n}\rangle$ and $|\Phi^{[\bf k]}_{\bf m}\rangle$ are provided by the expression

% {\color{blue}

\beq\label{densmat04}
 \langle \Phi^{[\bf k]}_{\bf n}|\rho_{[{\bf k}]}|\Phi^{[\bf k]}_{\bf m}\rangle := \rho_{[{\bf k}]}({\bf n}, {\bf m})  =  
\int_{\mathcal D}  d\mu ({\bf z},{\bar{\bf z}}) 
%{\color{blue}  \mathcal N(|{\bf z}|^{2})} 
P(|{\bf z}|^{2}) \,
\left\{
%\left(\mathcal N(|{\bf z}|^{2})\right)^{-1} 
\frac{{\bf z}^{\bf n}}{\sqrt{\left\{{\bf n}\right\}  !} }
\frac{\bar{\bf z}^{\bf m}}{\sqrt{\left\{{\bf m}\right\}  !} }\right\} 
\eeq
% }
which reduces to (\ref{densmat03}) at ${\bf m} = {\bf n}$. 

%{\color{blue}
%\brmk
Taking 

\beq
\rho_{[{\bf k}]} = \int_{\mathcal D} d\mu ({\bf z},{\bar{\bf z}})     
  P(|{\bf z}|^{2}) \, |\eta_{\bf z} \rangle \langle \eta_{\bf z} |
\eeq

instead of  the relation (\ref{densmat01}),  the expression (\ref{densmat03}) can be translated into 

\beq\label{densmat07}
\langle \Phi^{[\bf k]}_{\bf n}|\rho_{[{\bf k}]}|\Phi^{[\bf k]}_{\bf n}\rangle := \rho_{[{\bf k}]}({\bf n}, {\bf n}) = 
\int_{\mathcal D} d\mu ({\bf z},{\bar{\bf z}})  P(|{\bf z}|^{2}) \,\left(\mathcal N(|{\bf z}|^{2})\right)^{-1} 
\frac{|{\bf z}|^{2\bf n}}{\left\{{\bf n}\right\}  !}
\eeq

such that, by using (\ref{proba}), we obtain

\beq
\sum_{{\bf n} =0}^{\infty}  \langle \Phi^{[\bf k]}_{\bf n}|\rho_{[{\bf k}]}|\Phi^{[\bf k]}_{\bf n}\rangle  = 
\int_{\mathcal D} d\mu ({\bf z},{\bar{\bf z}})  P(|{\bf z}|^{2}) \,\left\{
\left(\mathcal N(|{\bf z}|^{2})\right)^{-1} 
\sum_{{\bf n} =0}^{\infty}  
\frac{|{\bf z}|^{2\bf n}}{\left\{{\bf n}\right\}  !}\right\} 
=1. 
\eeq

%\ermk
%}

From  the resolution of the identity
 and  (\ref{densmat05}), we derive the reproducing kernel property of the density matrix (see \cite{parthasarathy-sridhar}) 
by setting  $\langle \eta_{\bf z'} |\rho_{[{\bf k}]}|\eta_{\bf z} \rangle:= \rho_{[{\bf k}]}({\bf z'}, {\bf z})$, i.e.

\beq\label{kerndensity}
\rho_{[{\bf k}]}({\bf z'}, {\bf z}) &=& \langle \eta_{\bf z'} |
\int_{\mathcal D} d\mu ({\bf z''},{\bar{\bf z''}})|\eta_{\bf z''} \rangle \langle \eta_{\bf z''} |\rho_{[{\bf k}]}|\eta_{\bf z} \rangle\cr
&=& \int_{\mathcal D} d\mu ({\bf z''},{\bar{\bf z''}})\mathcal K({\bf z''},{\bf z'})\rho_{[{\bf k}]}({\bf z''}, {\bf z}).
\eeq

By making use of  the Proposition \ref{kernelprop}  and relation (\ref{kerndensity}), we infer here that the density matrix  also displays 
the self-reproducing kernel properties  as provided below:

\bpro 
\label{densmatkernel}

The following properties are satisfied by  $\rho_{[{\bf k}]}$ on the Hilbert space $\mathfrak H$:

\bitem

\item [(i)] Hermiticity  
\beq
\rho_{[{\bf k}]}({\bf z}, {\bf z'}) = \overline{\rho_{[{\bf k}]}({\bf z'}, {\bf z})}; 
\eeq

\item [(ii)] Positivity
\beq
\rho_{[{\bf k}]}({\bf z}, {\bf z})  >  0;
\eeq

\item [(iii)]  Idempotence

\beq\label{kern03}
\int_{\mathcal D}  d\mu ({\bf z''},{\bar{\bf z''}}) \rho_{[{\bf k}]}({\bf z}, {\bf z''})\rho_{[{\bf k}]}({\bf z''}, {\bf z'}) = 
\rho_{[{\bf k}]}({\bf z}, {\bf z'}).
\eeq

\eitem 

\epro

{\bf Proof.} See Appendix.

$\hfill{\square}$

Coming back to (\ref{matdensmat}), setting ${\bf z}^{\bf n}  := {\bf{J}}^{{\bf{n}}/2}e^{-\imath \varepsilon_{[\bf n]}\gamma}$, 
and  with

\beq
|\eta_{{\bf z}_{[{\bf k}]}}\rangle \langle \eta_{{\bf z}_{[{\bf k}]}}| = 
\left(\mathcal N(|{\bf z_{[{\bf k}]}}|^{2})\right)^{-1}
\sum_{[{\bf{m}}], [{\bf{n}}]=0}^{\infty}\frac{{\bf z}^{\bf n}_{[{\bf k}]}}{\sqrt{\left\{{\bf n}\right\} !} }
\frac{{\bar{\bf z}}^{\bf m}_{[{\bf k}]}}{\sqrt{\left\{{\bf m}\right\} !} }
|\Psi_{[\bf k]}\otimes\Phi^{[\bf k]}_{[{\bf{n}}]}\rangle \langle \Psi_{[\bf k]}\otimes\Phi^{[\bf k]}_{[{\bf{m}}]}|
\eeq

%and  
$d\mu(\mathfrak Z, \bar{\mathfrak Z})  =    
d\mu(\gamma)d\nu(\mathfrak J)$ where $d\mu(\gamma) = \prod_{[\bf k] \in \Gamma}d\mu(\gamma_{[\bf k]})$, 
$d\nu(\mathfrak J) = \prod_{[\bf k] \in \Gamma}d\nu({\bf J}_{[\bf k]}),$
and taking  $\mathcal D =  ([0,\infty) \times [0,2\pi))^{N_B}$, we get

\beq
\hat \rho &=&  \sum_{[{\bf k}]\in \Gamma}\sum_{{[\bf j],[\bf q]}=0}^{\infty}
\int_{\mathcal D}d\mu(\gamma_{(0,\cdots,0,0)})d\nu_{(0,\cdots,0,0)}({\bf J}_{(0,\cdots,0,0)})
\int_{\mathcal D}d\mu(\gamma_{(1,\cdots,0,0)})d\nu_{(1,\cdots,0,0)}({\bf J}_{(1,\cdots,0,0)})\cdots \crcr
&& \int_{\mathcal D}d\mu(\gamma_{[{\bf k}]})d\nu_{[{\bf k}]}({\bf J}_{[{\bf k}]})\cdots 
\int_{\mathcal D}d\mu(\gamma_{(1,\cdots,1,1)})d\nu_{(1,\cdots,1,1)}({\bf J}_{(1,\cdots,1,1)})\times \crcr
&& diag\left(\frac{{\bf J}^{({\bf j}+{\bf q})/2}_{(0,0, \dots, 0,0)}
e^{-\imath (\varepsilon_{[\bf j]} - \varepsilon_{[\bf q]})\gamma_{(0,0, \dots, 0,0)}}
 P({\bf J}_{(0,0, \dots, 0,0)})}{\sqrt{\{{\bf j}\}! \{{\bf q\}!} }} , 
\frac{{\bf J}^{({\bf j}+{\bf q})/2}_{(1,0, \dots 0,0)}
e^{-\imath (\varepsilon_{[\bf j]} - \varepsilon_{[\bf q]})\gamma_{(1,0, \dots, 0,0)}}P({\bf J}_{(1,0, \dots 0,0)})}
{\sqrt{\{{\bf j}\}! \{{\bf q\}!} }}
 \right.\nonumber
\\ 
&&  \left. \cdots,  \frac{{\bf J}^{({\bf j}+{\bf q})/2}_{[{\bf k}]}
e^{-\imath (\varepsilon_{[\bf j]} - \varepsilon_{[\bf q]})\gamma_{[\bf k]}}P({\bf J_{[{\bf k}]}})}{\sqrt{\{{\bf j}\}! \{{\bf q\}!} }}, 
\cdots, \frac{{\bf J}^{({\bf j}+{\bf q})/2}_{(1,1, \dots, 1,1)}
e^{-\imath (\varepsilon_{[\bf j]} - \varepsilon_{[\bf q]})\gamma_{(1,1, \dots, 1,1)}}P({\bf J}_{(1,1, \dots, 1,1)})}
{\sqrt{\{{\bf j}\}! \{{\bf q\}!} }} 
  \right)\cr
&&\times
 |\Psi_{[\bf k]}
\otimes\Phi^{[\bf k]}_{[{\bf{j}}]}\rangle
\langle\Psi_{[\bf k]}\otimes\Phi^{[\bf k]}_{[{\bf{q}}]}|. 
\eeq

Therefore, the analogue to (\ref{densmat04}) is given by

\beq
&&\langle\Psi_{[\bf k]}\otimes\Phi^{[\bf k]}_{[{\bf{n}}]}|\hat \rho
|\Psi_{[\bf k]}
\otimes\Phi^{[\bf k]}_{[{\bf{p}}]}\rangle \cr
&&=  
  (2\pi)^{N_B}  \int_{0}^{\infty} \int_{0}^{\infty}\cdots \int_{0}^{\infty} d\nu_{(0,\cdots,0,0)}({\bf J}_{(0,\cdots,0,0)})
\int_{0}^{\infty} \int_{0}^{\infty}\cdots \int_{0}^{\infty}d\nu_{(1,\cdots,0,0)}({\bf J}_{(1,\cdots,0,0)})\cdots \crcr
&& \int_{0}^{\infty} \int_{0}^{\infty}\cdots \int_{0}^{\infty} d\nu_{[{\bf k}]}({\bf J}_{[{\bf k}]})\cdots 
\int_{0}^{\infty} \int_{0}^{\infty}\cdots \int_{0}^{\infty} d\nu_{(1,\cdots,1,1)}({\bf J}_{(1,\cdots,1,1)})\times \crcr
&& diag \left(\frac{{\bf J}^{({\bf n}+{\bf p})/2}_{(0,0, \dots, 0,0)}
}{\sqrt{\{{\bf n}\}! \{{\bf p\}!}}} 
 P({\bf J}_{(0,0, \dots, 0,0)}), 
\frac{{\bf J}^{({\bf n}+{\bf p})/2}_{(1,0, \dots 0,0)}
}{\sqrt{\{{\bf n}\}! \{{\bf p\}!}}} P({\bf J}_{(1,0, \dots 0,0)}) \cdots,  \frac{{\bf J}^{({\bf n}+{\bf p})/2}_{[{\bf k}]}
}{\sqrt{\{{\bf n}\}! \{{\bf p\}!}}}  P({\bf J_{[{\bf k}]}}), 
\cdots, \right.
\crcr
&&  \left.\frac{{\bf J}^{({\bf n}+{\bf p})/2}_{(1,1, \dots, 1,1)}
}{\sqrt{\{{\bf n}\}! \{{\bf p\}!}}} P({\bf J}_{(1,1, \dots, 1,1)}) \right)\, \delta_{{[\bf n],[\bf p]}} 
\eeq

where the following  identity 

\beq
\int_{0}^{2\pi}\int_{0}^{2\pi}\cdots \int_{0}^{2\pi} 
d\mu(\gamma_{[{\bf k}]})e^{-\imath (\varepsilon_{[\bf n]} - \varepsilon_{[\bf p]})\gamma_{[\bf k]}} = 
\left\{
              \begin{array}{lll}
              0  \quad \mbox{if}  \quad [\bf n] \neq [\bf p], \\
               \\
              (2\pi)^{N_B}     
\quad \mbox{if}  \quad [\bf n] = [\bf p]
               \end{array}
\right.
\eeq
is used.

\section{NCS quantization of the complex plane} 
Provided  the resolution of the identity satisfied by the NCS,  in this section we perform the NCS quantization of   phase space  classical 
 observables.   For more details in the CS quantization procedure, see \cite{gazbook09}.
 As a matter of illustration, the case of $q$-deformed CS is explicitly treated.

\subsection{General construction}
 Such a quantization is realized through the mappings:

\begin{itemize}
\item [(i)]
\beq\label{quantiz00}
A_{\bf z}:  \mathcal D \rightarrow  \mathfrak H, \quad\quad
%\cr
{\bf z} &\mapsto&  \int_{\mathcal D}{\bf z} \, |\eta_{\bf z}\rangle \langle \eta_{\bf z}|
\mathcal N(|{\bf z}|^{2})d\mu ({\bf z},{\bar{\bf z}})
 \cr
&=& \sum_{{\bf n}=0}^{\infty}\sqrt{({\bf n}+1)  }f({\bf n}+1) 
|\Phi^{[\bf k]}_{\bf n}\rangle \langle \Phi^{[\bf k]}_{{\bf n}+1}|, 
\eeq

\item [(ii)]
\beq\label{quantiz01}
A_{\bar{\bf z}}:  \mathcal D \rightarrow  \mathfrak H,\quad\quad 
{\bf z} &\mapsto&   \int_{\mathcal D}\bar{\bf z} \, |\eta_{\bf z}\rangle \langle \eta_{\bf z}|
\mathcal N(|{\bf z}|^{2})d\mu ({\bf z},{\bar{\bf z}}) \cr
&=& \sum_{{\bf n}=0}^{\infty}\sqrt{{\bf n}  }f({\bf n}) 
|\Phi^{[\bf k]}_{\bf n}\rangle \langle \Phi^{[\bf k]}_{{\bf n}-1}|,
\eeq

\item [(iii)]
\beq
A'_{\bf z} :  \tilde{\mathcal D} \rightarrow  \mathfrak H, \quad\quad
{\bf z} &\mapsto&  \int_{\tilde{\mathcal D}}{\bf z} \, |\eta'_{\bf z}\rangle \langle \eta'_{\bf z}|
\mathcal N'(|{\bf z}|^{2})d\mu ({\bf z},{\bar{\bf z}}) \cr
 &=& \sum_{{\bf n}=0}^{\infty}\frac{\sqrt{({\bf n}+1)  }}{f({\bf n}+1) }
|\Phi^{[\bf k]}_{\bf n}\rangle \langle \Phi^{[\bf k]}_{{\bf n}+1}|, 
\eeq

\item [(iv)]
\beq
A'_{\bar{\bf z}} :   \tilde{\mathcal D}  \rightarrow  \mathfrak H, \quad\quad
{\bf z} &\mapsto& \int_{\tilde{\mathcal D}}\bar{\bf z} \, |\eta'_{\bf z}\rangle \langle \eta'_{\bf z}|
\mathcal N'(|{\bf z}|^{2})d\mu ({\bf z},{\bar{\bf z}}) \cr
&=&\sum_{{\bf n}=0}^{\infty}\frac{\sqrt{{\bf n}  }}{f({\bf n}) }
|\Phi^{[\bf k]}_{\bf n}\rangle \langle \Phi^{[\bf k]}_{{\bf n}-1}|.
\eeq

\end{itemize}

The physical utility of NCS in different applications consists in the calculation of the expectation (mean) 
values of quantized elementary classical observables. 
In this spirit,  the  following mean values are computed: 

\beq\label{quant00}
\langle \eta_{\bf z}| A_{\bf z}|\eta_{\bf z}\rangle = {\bf z}, \qquad  \langle \eta_{\bf z}|  A_{\bar{\bf z}} |\eta_{\bar{\bf z}}\rangle 
= \bar{\bf z},\qquad 
%\eeq
%\beq\label{quant01}
\langle \eta_{\bf z}|A^{2}_{\bf z}|\eta_{\bf z}\rangle = {\bf z}^2, \qquad  
\langle \eta_{\bf z}|A^{2}_{\bar{\bf z}}|\eta_{\bf z}\rangle = {\bar{\bf z}}^2,
\eeq

 \beq\label{quant02}
\langle \eta_{\bf z}|A_{\bar{\bf z}}A_{\bf z}|\eta_{\bf z}\rangle &=&  |{\bf z}|^2 
\eeq

\beq\label{quant03}
\langle \eta_{\bf z}|A_{\bf z}A_{\bar{\bf z}}|\eta_{\bf z}\rangle &=& \left(\mathcal N(|{\bf z}|^{2})\right)^{-1}\cr 
&& \times \left[ |{\bf z}|^2 \sum_{{\bf n}=1}^{\infty} 
\frac{[f({\bf n}+1) ]^2 }{[f({\bf n}) ]^2}
\frac{|{\bf z}|^{2({\bf n}-1)}}{({\bf n}-1) ![f({\bf n}-1) !]^2} + \sum_{{\bf n}=0}^{\infty}
\frac{[f({\bf n}+1) ]^2}{[f({\bf n}) !]^2} 
\frac{|{\bf z}|^{2{\bf n}}}{{\bf n} !}\right]
\nonumber \\
\eeq

such that for   $f({\bf n}) \rightarrow 1$ with $f({\bf n}+1)= 1 = f({\bf n}), \, f({\bf n}+1) != 1 = f({\bf n}) !,$
%for $f({\bf n}) \rightarrow 1$ where  $f({\bf n}+1)= 1 = f({\bf n}), \, f({\bf n}+1) != 1 = f({\bf n}) !$
we get 

\beq
\langle \eta_{\bf z}|A_{\bf z}A_{\bar{\bf z}}|\eta_{\bf z}\rangle = |{\bf z}|^2  + \langle \eta_{\bf z}|\mathbb I|\eta_{\bf z}\rangle
\eeq
as in the case of the standard CS.  
Moreover, from  the following computed quantities:

\beq
A_{|{\bf z}|^2} &=& \sum_{{\bf n}=0}^{\infty} ({\bf n}+1)  [f({\bf n}+1) ]^2
|\Phi^{[\bf k]}_{\bf n}\rangle \langle \Phi^{[\bf k]}_{{\bf n}}|, \cr
\cr
A_{{\bf z}^2} &=& \sum_{{\bf n}=0}^{\infty}\sqrt{({\bf n}+2)({\bf n}+1)}f({\bf n} +2)f({\bf n} +1)
|\Phi^{[\bf k]}_{\bf n}\rangle \langle \Phi^{[\bf k]}_{{\bf n}+2}|, \cr
\cr
A_{\bar{\bf z}^2} &=& \sum_{{\bf n}=0}^{\infty}\sqrt{({\bf n}+2)({\bf n}+1)}f({\bf n} +2)f({\bf n} +1)
|\Phi^{[\bf k]}_{{\bf n}+2}\rangle \langle \Phi^{[\bf k]}_{{\bf n}}|,
\eeq

 we derive the  commutators: 
%between   $A_{|{\bf z}|^2}$ and $ A_{{\bf z}}, A_{\bar{\bf z}}, A_{{\bf z}^2}, A_{\bar{\bf z}^2}$ as follows:

\bitem
%{\color{blue}
\item

\beq
[A_{|{\bf z}|^2}, A_{{\bf z}}]  &=&  
\sum_{{\bf n}=0}^{\infty} \sqrt{({\bf n}+1)  }f({\bf n}+1)\cr
&& \times \left\{
({\bf n}+1)  [f({\bf n}+1) ]^2 - ({\bf n}+2)  [f({\bf n}+2) ]^2 \right\} 
|\Phi^{[\bf k]}_{\bf n}\rangle \langle \Phi^{[\bf k]}_{{\bf n}+1}|;
\eeq
\item

\beq
[A_{|{\bf z}|^2}, A_{\bar{\bf z}}] &=&  \sum_{{\bf n}=0}^{\infty} \sqrt{({\bf n}+1)  }f({\bf n}+1) \cr
&&\times\left\{
({\bf n}+2)  [f({\bf n}+2) ]^2 - ({\bf n}+1)  [f({\bf n}+1) ]^2 \right\}
|\Phi^{[\bf k]}_{{\bf n}+1}\rangle \langle \Phi^{[\bf k]}_{{\bf n}}|; 
\eeq
%}

\item

\beq
[A_{|{\bf z}|^2}, A_{{\bf z}^2}] &=&   \sqrt{({\bf n}+2)({\bf n}+1)}f({\bf n} +2)f({\bf n} +1) \cr
&& \times \left\{({\bf n}+1)  [f({\bf n}+1) ]^2  - ({\bf n}+3)  [f({\bf n}+3) ]^2 \right\}
|\Phi^{[\bf k]}_{\bf n}\rangle \langle \Phi^{[\bf k]}_{{\bf n}+2}|; \nonumber \\
\eeq

\item

\beq
[A_{|{\bf z}|^2}, A_{\bar{\bf z}^2}] &=&   \sqrt{({\bf n}+2)({\bf n}+1)}f({\bf n} +2)f({\bf n} +1) \cr
&& \times \left\{ ({\bf n}+3)  [f({\bf n}+3) ]^2 - ({\bf n}+1)  [f({\bf n}+1) ]^2 \right\}
|\Phi^{[\bf k]}_{{\bf n}+2}\rangle  \langle \Phi^{[\bf k]}_{{\bf n}}|. \nonumber \\
\eeq

\eitem

Considering the usual phase space conjugate coordinates $({\bf q},{\bf p})$ through ${\bf z} 
= \frac{{\bf q} +i{\bf p}}{\sqrt{2}},$ we get for the classical
{\it position}    function ${\bf q}$ : 

\beq\label{op00}
Q :=  A_{{\bf q}} = \frac{1}{\sqrt{2}}(A_{\bf z} + A_{\bar{\bf z}}),
\eeq

and for the classical {\it momentum}     function ${\bf p}$ : 

\beq\label{op01}
P :=  A_{{\bf p}} = \frac{1}{i\sqrt{2}}(A_{\bf z} -  A_{\bar{\bf z}}).
\eeq

The mean values of the operators $Q, P$ and $Q^2, P^2$ can be written as follows:

\beq
\langle \eta_{\bf z}|Q |\eta_{\bf z}\rangle &=& \frac{1}{\sqrt{2}}[\langle \eta_{\bf z}| A_{\bf z}|\eta_{\bf z}\rangle +  
\langle \eta_{\bf z}|  A_{\bar{\bf z}} |\eta_{\bf z}\rangle] = 
\frac{1}{\sqrt{2}}({\bf z} + \bar{\bf z}) = {\bf q}, \cr
\langle \eta_{\bf z}|P |\eta_{\bf z}\rangle   
&=&  \frac{1}{i\sqrt{2}}[\langle \eta_{\bf z}| A_{\bf z}|\eta_{\bf z}\rangle - \langle \eta_{\bf z}|  A_{\bar{\bf z}} |\eta_{\bf z}\rangle] = 
\frac{1}{i\sqrt{2}}({\bf z} - \bar{\bf z}) = {\bf p}
\eeq

\beq\label{quant04}
\langle \eta_{\bf z}|Q^2 |\eta_{\bf z}\rangle 
&=& \frac{1}{2} \left[\langle \eta_{\bf z}|A^{2}_{\bf z}|\eta_{\bf z}\rangle + \langle \eta_{\bf z}|A^{2}_{\bar{\bf z}}|\eta_{\bf z}\rangle + 
\langle \eta_{\bf z}|\left\{A_{\bf z}, A_{\bar{\bf z}}\right\}|\eta_{\bf z}\rangle   \right]  \cr
\langle \eta_{\bf z}|P^2 |\eta_{\bf z}\rangle &=& 
-\frac{1}{2} \left[\langle \eta_{\bf z}|A^{2}_{\bf z}|\eta_{\bf z}\rangle + \langle \eta_{\bf z}|A^{2}_{\bar{\bf z}}|\eta_{\bf z}\rangle - 
\langle \eta_{\bf z}|\left\{A_{\bf z}, A_{\bar{\bf z}}\right\}|\eta_{\bf z}\rangle  \right]
\eeq

with the anti-commutator of two operators $\mathcal A$ and $\mathcal B$   given by $\left\{\mathcal A, \mathcal B\right\} := 
\mathcal A \mathcal B + \mathcal B \mathcal A$, where

\beq\label{quant05}
\langle \eta_{\bf z}|\left\{A_{\bf z}, A_{\bar{\bf z}}\right\}|\eta_{\bf z}\rangle &=& 
\left(\mathcal N(|{\bf z}|^{2})\right)^{-1}\cr
&& \times \left[\sum_{{\bf m} =0}^{\infty}
\left\{({\bf m}+1)[f({\bf m}+1)]^2 +  {\bf m}[f({\bf m})]^2 \right\}\times \frac{|{\bf z}|^{2{\bf m}}}{\left\{{\bf m}\right\}!}\right].
\eeq  
  
Using (\ref{quant00})-(\ref{quant03}), expressions (\ref{quant04}) are obtained as 

\beq
\langle \eta_{\bf z}|Q^2 |\eta_{\bf z}\rangle 
&=& \frac{1}{2} \left[{\bf q}^2-{\bf p}^2 + 
\langle \eta_{\bf z}|\left\{A_{\bf z}, A_{\bar{\bf z}}\right\}|\eta_{\bf z}\rangle   \right]
\eeq
\beq
\langle \eta_{\bf z}|P^2 |\eta_{\bf z}\rangle 
&=& -\frac{1}{2} \left[{\bf q}^2-{\bf p}^2 - 
\langle \eta_{\bf z}|\left\{A_{\bf z}, A_{\bar{\bf z}}\right\}|\eta_{\bf z}\rangle   \right]  
\eeq
with $\langle \eta_{\bf z}|\left\{A_{\bf z}, A_{\bar{\bf z}}\right\}|\eta_{\bf z}\rangle$ given in (\ref{quant05}).

Then,   with $\left\{{\bf m}  \right\} :=  {\bf m}[f({\bf m})]^2 $, we get the  dispersions:

\beq\label{quad000}
(\Delta Q)^2 &=& \langle Q^2\rangle -  (\langle Q\rangle)^2 =
 -|{\bf z}|^{2}  + \frac{1}{2}\left(\mathcal N(|{\bf z}|^{2})\right)^{-1}\left[\sum_{{\bf m} =0}^{\infty}
\left[\left\{{\bf m}+1\right\} +  \left\{{\bf m}  \right\}\right]\times \frac{|{\bf z}|^{2{\bf m}}}{\left\{{\bf m}\right\}!}\right]\cr
&=& -\frac{1}{2}  |{\bf z}|^{2}  +  \mathcal G(|{\bf z}|^{2})
\eeq

\beq
(\Delta P)^2 &=& \langle P^2\rangle -  (\langle P\rangle)^2 
= -|{\bf z}|^{2}  + \frac{1}{2}\left(\mathcal N(|{\bf z}|^{2})\right)^{-1} \left[\sum_{{\bf m} =0}^{\infty}
\left[\left\{{\bf m}+1\right\} +  \left\{{\bf m}  \right\}\right]\times \frac{|{\bf z}|^{2{\bf m}}}{\left\{{\bf m}\right\}!}\right]\cr
&=& -\frac{1}{2}  |{\bf z}|^{2}  +  \mathcal G(|{\bf z}|^{2}) 
\eeq

where 

\beq\label{quad001}
\mathcal G(|{\bf z}|^{2}) :=  \frac{1}{2}\left(\mathcal N(|{\bf z}|^{2})\right)^{-1} 
\left[\sum_{{\bf m} =0}^{\infty}
\left\{{\bf m}+1 \right\}\times \frac{|{\bf z}|^{2{\bf m}}}{\left\{{\bf m}\right\}!}\right].
\eeq

%{\color{blue}

Now, taking into account both (\ref{op00}) and (\ref{op01}), and by defining $   
\langle   [Q, P]  \rangle :=  \langle \eta_{\bf z}|[Q, P]| \eta_{\bf z}\rangle$,
we get

\beq\label{quad002}
\langle   [Q, P]  \rangle = i 
\left(\mathcal N(|{\bf z}|^{2})\right)^{-1}\left[\sum_{{\bf m} =0}^{\infty}
\left[\left\{{\bf m}+1\right\} - \left\{{\bf m}  \right\}\right]\times \frac{|{\bf z}|^{2{\bf m}}}{\left\{{\bf m}\right\}!}\right]
\eeq

which leads, by using (\ref{quad000})-(\ref{quad001}),  to 

\beq\label{quad003}
(\Delta Q)^2 = (\Delta P)^2 = \frac{1}{2}\left|\langle   [Q, P]  \rangle \right|
\eeq

and, therefore, to the relation

\beq\label{quadsat}
(\Delta Q)^2  (\Delta P)^2 = \frac{1}{4} \left|\langle   [Q, P]  \rangle \right|^2
\eeq

attesting that the NCS $| \eta_{\bf z}\rangle $ are intelligent, as in the standard situation, 
 for the operators  $Q $ and $P$ obtained from the CS quantization  
of the classical observables ${\bf q}$ and ${\bf p}$, respectively.

%}

\subsection{Illustration: case of Quesne's $q$-deformed CS \cite{quesne}}
The Quesne's $q$-deformed CS \cite{quesne} can be retrieved from the above generalization by setting 
 $f({\bf n})= \sqrt{\frac{[{\bf n}]_q}{{\bf n}}}$, 
with $0 < q < 1$ and ${\bf z} \in \C,$  and

\beq
|\eta_{\bf z}\rangle_q = \left(\mathcal N_q(|{\bf z}|^{2})\right)^{-1/2}\sum_{ {\bf n}=0}^{\infty}
  \frac{{\bf z}^{{\bf n}}}{\sqrt{[{\bf n}]_q !} }
|\Phi^{[\bf k]}_{{\bf n}}\rangle_q, \qquad  \mathcal N_q(|{\bf z}|^{2}) = \sum_{{\bf n} =0}^{\infty}\frac{|{\bf z}|^{2{\bf n}}}{[{\bf n}]_q !} = 
E_q(|{\bf z}|^{2})
\eeq
 
where the $q$-factorial

\beq
[{\bf n}]_q !  \equiv  \left\{
              \begin{array}{lll}
              1    \quad \quad \quad \quad \quad \quad \quad \quad  \mbox{if}  \quad  {\bf n} = 0 \\
\\
              {[\bf n]}_q [{\bf n}-1]_q \dots [1]_q  \quad \mbox{if} \quad {\bf n} = 1, 2, 3, \dots 
                \end{array}
\right. 
\eeq

is defined in terms of the $q$-numbers
\beq\label{qnumber}
[{\bf n}]_q \equiv \frac{q^{{\bf n}} -1}{q-1} = 1 + q +q^2 + \dots + q^{{\bf n}-1} 
\eeq
such that in the limit $q\rightarrow 1$, $[{\bf n}]_q, [{\bf n}]_q !$ tend to ${\bf n}$ and ${\bf n} !$, respectively.

 In this case the deformation (\ref{eq1}), with $f_l(N_l) = \sqrt{\frac{[N_l]_q}{N_l}}$, 
  is obtained through the following correspondences:

\beq\label{qnumb00}  
a_l \rightarrow b_l =  a_l \sqrt{\frac{[N_l]_q}{N_l}},  \qquad 
 a^{\dag}_{l} \rightarrow  b^{\dag}_{l} = \sqrt{\frac{[N_l]_q}{N_l}} a^{\dag}_l,  \qquad 1\leq l \leq N_B
\eeq

where $N_l =  a^{\dag}_{l} a_l$ and $[N_l]_q$ defined as in (\ref{qnumber}). From (\ref{qnumb00}), it comes that  $b^{\dag}_{l}b_l = 
[N_l]_q$ and $b_l  b^{\dag}_{l} = [N_l  + 1]_q$. The actions of the operators $b_l,   b^{\dag}_{l}$ on the states 
$|\Phi^{[\bf k]}_{{n}_l}\rangle_q$ are   
given for $1\leq l \leq N_B$ by 

\beq
b_l|\Phi^{[\bf k]}_{{n}_l}\rangle_q = \sqrt{[n_l]_q}|\Phi^{[\bf k]}_{{n}_l-1}\rangle_q, \quad  
b^{\dag}_{l}|\Phi^{[\bf k]}_{{n}_l}\rangle_q = \sqrt{[n_l +1]_q}|\Phi^{[\bf k]}_{{n}_l+1}\rangle_q,  \quad  
b_lb^{\dag}_{l}-q b^{\dag}_{l}b_l= \mathbb I 
\eeq
where in the limit $q\rightarrow 1$, $[b_l, b^{\dag}_{l}] = \mathbb I$. 

The   states   $|\Phi^{[\bf k]}_{{\bf n}}\rangle_q = \bigotimes_{l=1}^{N_B}|\Phi^{[\bf k]}_{n_{l}}\rangle_q $  are such that 
the states $|\Phi^{[\bf k]}_{n_{l}}\rangle_q$ are obtained with  (\ref{eigvec}) as

\beq
|\Phi^{[\bf k]}_{n_{l}}\rangle_q 
%= e^{i \sqrt{2}\frac{g_{[\bf k]}}{\omega_{l}}\mathcal P_{l}}|\phi_{n_{l}}\rangle 
= \frac{(\mathcal A^{\dag}_{[{\bf k}]_{l}})^{n_{l}}}{\sqrt{[n_{l}]_q !}  }|\Phi^{[\bf k]}_{0_{l}}\rangle_q,  \quad 
\mathcal A_{[{\bf k}]_{l}} = 
e^{i \sqrt{2} {\frac{g_{[\bf k]}}{\omega_{l}}}\mathcal P_{l}}b_{l}e^{-i\sqrt{2}\frac{g_{[\bf k]}}{\omega_{l}}\mathcal P_{l}}
\eeq

with  $|\Phi^{[\bf k]}_{0_{l}}\rangle_q  = e^{i \sqrt{2} \frac{g_{[\bf k]}}{\omega_{l}}\mathcal P_{l}}|0\rangle_q$,  where we assume 
$|0\rangle_q = |0\rangle,  (b_l|0\rangle_q=0)$, and $\mathcal P_{l}$ with $f_l(N_l) = \sqrt{\frac{[N_l]_q}{N_l}}$ is given as in (\ref{moment00}). 
The $|\Phi^{[\bf k]}_{{\bf n}}\rangle_q$ are identified with the states $|n\rangle_q$ of the Hilbert space $\mathcal F_q$  \cite{quesne} which is 
 the $q$-deformed Fock space associated with the $q$-boson creation and annihilation
operators.

Provided the resolution of the identity on  $\mathcal F_q$ spanned by the states $|\Phi^{[\bf k]}_{{\bf n}}\rangle_q$
 
\beq
\int \int_{\C} d^{2}{\bf z} |\eta_{\bf z}\rangle_q \, W_q(|{\bf z}|^{2})\,  _q\langle \eta_{\bf z}|  = \sum_{ {\bf n}=0}^{\infty}
|\Phi^{[\bf k]}_{{\bf n}}\rangle_q \, _q\langle \Phi^{[\bf k]}_{{\bf n}}| = I_{\mathcal F_q}
\eeq
where 

\beq
W_q(|{\bf z}|^{2}) = \frac{q-1}{\pi \ln q }\frac{E_q(|{\bf z}|^{2})}{E_q(q|{\bf z}|^{2})},
\eeq

the CS quantization is performed as in (\ref{quantiz00})-(\ref{quantiz01}) and leads to the $q$-deformed analogue of  (\ref{quant05}):

%{\color{blue}  (verify)
\beq 
_q\langle \eta_{\bf z}|\left\{A_{\bf z}, A_{\bar{\bf z}}\right\}|\eta_{\bf z}\rangle_q &=& 
\left(\mathcal N_q(|{\bf z}|^{2})\right)^{-1} \left[\sum_{{\bf m} =0}^{\infty}
\left\{[{\bf m}+1]_q  +  [{\bf m}]_q  \right\}\times \frac{|{\bf z}|^{2{\bf m}}}{[{\bf m}]_q!}\right].
\eeq 

The dispersions are given by  

\beq\label{quad004}
(\Delta Q)_q^2 = (\Delta P)_q^2 = -\frac{1}{2}|{\bf z}|^{2} + \mathcal G_q(|{\bf z}|^{2})   
\eeq

where

\beq\label{quad005}
\mathcal G_q(|{\bf z}|^{2})  := \frac{1}{2}\left(\mathcal N_q(|{\bf z}|^{2})\right)^{-1} 
\left[\sum_{{\bf m} =0}^{\infty}
[{\bf m}+1]_q  \frac{|{\bf z}|^{2{\bf m}}}{[{\bf m}]_q !}\right].
\eeq

%}
%{\color{blue} 

Proceeding as in (\ref{quad000})-(\ref{quad003}) with (\ref{quad004}) and (\ref{quad005}), we arrive at

\beq
(\Delta Q)^2_q  (\Delta P)^2_q = \frac{1}{4} \left|\langle [Q, P]_q\rangle \right|^2.
\eeq

%with the same conclusion as for (\ref{quadsat}).

%}

%{\color{blue} Give the $q$-deformed expression of $\mathcal G({\bf z})$ for $f({\bf n})= \sqrt{\frac{[{\bf n}]_q}{{\bf n}}}$}

\section{Quantum optical features of the NCS}

Now, we exploit the results issued from the above developed quantization procedure  
to inspect   some quantum optical properties of the
constructed NCS. Roy {\it et  al} \cite{roy-roy}   pointed out  
interesting properties exhibited by NCS, such as squeezing and sub-Poissonian behaviour. For more details see also \cite{dodonov}. 
In the following, we  discuss   some other relevant aspects 
 like  the signal-to-quantum-noise ratio and the Mandel parameter.

\subsection{Signal-to-quantum-noise ratio (SNR)}

Signal-to-quantum-noise ratio (SNR) is relevant    
when  studying, for example, the
exciton spin relaxation for dynamics of photoexcited excitons in
an ensemble of InAs/GaAs self-assembled quantum dots \cite{mulleretal}. For a normalized state $|\phi \rangle$, in terms of 
the self-adjoint 
quadrature operator $Q$, the SNR is defined as 
 \cite{penson-solomon} 

\beq
\sigma_{|\phi \rangle} = \frac{\langle Q \rangle^{2}_{|\phi \rangle}}{(\Delta Q)^{2}_{|\phi \rangle}}.
\eeq

In the NCS $|\eta_{\bf z}\rangle  $, we obtain  

\beq
\sigma_{\bf z} =   \frac{{\bf q}^2}{(\Delta Q)^2  } 
=    \frac{2|{\bf z}|^2 \cos^2{\phi}}{-\frac{1}{2}  |{\bf z}|^{2}    +  \mathcal G(|{\bf z}|^{2})}.
\eeq
 
In the case where $f({\bf n}) \rightarrow 1$ with $f({\bf n}+1)= 1 = f({\bf n}), \, f({\bf n}+1) != 1 = f({\bf n}) !$, we have 
$\mathcal G(|{\bf z}|^{2}) = \frac{1}{2} + \frac{1}{2}  |{\bf z}|^{2} $ providing $\sigma_{\bf z} = 4|{\bf z}|^2 \cos^2{\phi}$, this latter 
being the SNR for the canonical CS (CCS).

For the $q$-deformed CS, the SNR is given by  

\beq
\sigma_{{\bf z}_q} =   \frac{{\bf q}^2}{(\Delta Q)^2_q  } 
=    \frac{2|{\bf z}|^2 \cos^2{\phi}}{-\frac{1}{2}  |{\bf z}|^{2}    +  \mathcal G_q(|{\bf z}|^{2})} 
\eeq
such that in the limit $q \rightarrow 1$, $\mathcal G_q(|{\bf z}|^{2})  \rightarrow \frac{1}{2} + \frac{1}{2}  |{\bf z}|^{2}$ leading to 
$\sigma_{{\bf z}_q} \rightarrow 4|{\bf z}|^2 \cos^2{\phi}$.

\subsection{Mandel parameter}

The Mandel parameter   known as a convenient noise-indicator of a non-classical field  and defined by \cite{mandel-wolf}

\beq
\mathcal Q = \frac{(\Delta N)^{2}}{\langle N \rangle} - 1 \equiv \mathcal F - 1
\eeq

 is closely related to the normalized variance  also called the quantum Fano factor $\mathcal F$ \cite{bajer-miranowicz}, 
  given   by $\mathcal F = (\Delta N)^{2}/\langle N \rangle$, of the photon distribution. For 
$\mathcal F < 1 (\mathcal Q \leq 0)$, the emitted light 
  is referred to as sub-Poissonian, with $(\mathcal F = 1; \mathcal Q = 0)$, whereas for $\mathcal F > 1, (\mathcal Q > 0)$ the 
  light is called super-Poissonian.

%{\color{blue}
We get for $\mathcal N = A^{\dag} A $,  where $A$ and $A^{\dag}$ are given as in (\ref{opexp}) with 
$A|\Phi^{[\bf k]}_{\bf n}\rangle = \sqrt{\left\{{\bf n}\right\}  }
|\Phi^{[\bf k]}_{\bf n - 1}\rangle$ and $A^{\dag}|\Phi^{[\bf k]}_{\bf n}\rangle = \sqrt{\left\{{\bf n} +1\right\}  }
|\Phi^{[\bf k]}_{\bf n + 1}\rangle$,  the following mean values
%}
 
\beq
\langle \eta_{\bf z}|\mathcal N|\eta_{\bf z}\rangle 
&=&   
 |{\bf z}|^{2}
\left(\mathcal N(|{\bf z}|^2)\right)^{-1}     \sum_{{\bf n}=1}^{\infty}
\frac{|{\bf z}|^{2({\bf n}-1)}}{\left\{({\bf n}-1)\right\} !} = |{\bf z}|^{2}
\eeq

and

\beq
\langle \eta_{\bf z}|\mathcal N^2|\eta_{\bf z}\rangle
 = \left(\mathcal N(|{\bf z}|^2)\right)^{-1} \sum_{{\bf n}=0}^{\infty}
\frac{|{\bf z}|^{2{\bf n}}}{ \left\{{\bf n}\right\} !}\left\{{\bf n}\right\}^2.
\eeq

Thus, the dispersion is derived in the NCS $|\eta_{\bf z}\rangle  $  as follows:

\beq
(\Delta \mathcal N)^2_{\bf z} &=&  \langle \mathcal N^2 \rangle_{\bf z}   - (\langle \mathcal N \rangle_{\bf z})^2  \cr
&=& \left(\mathcal N(|{\bf z}|^2)\right)^{-1} \sum_{{\bf n}=0}^{\infty}
\frac{|{\bf z}|^{2{\bf n}}}{ \left\{{\bf n}\right\} !}\left\{{\bf n}\right\}^2
 -  (|{\bf z}|^2)^2.   
\eeq

The Mandel parameter is therefore provided for a given $f$-deformed function of the number operator   in the manner

\beq
\mathcal Q &=& \frac{(\Delta \mathcal N)^{2}_{\bf z}}{\langle \mathcal  N \rangle_{\bf z}} - 1  \cr
\cr
&=& \left(\mathcal N(|{\bf z}|^2)\right)^{-1} \sum_{{\bf n}=0}^{\infty}
\frac{|{\bf z}|^{2{\bf n}}}{ \left\{{\bf n}\right\} !}\left\{{\bf n}+1\right\}  - (|{\bf z}|^2 + 1).
\eeq

In the case where $f({\bf n}) \rightarrow 1$ with $f({\bf n}+1)= 1 = f({\bf n}), \, f({\bf n}+1) != 1 = f({\bf n}) !$, we have the quantities related 
to the canonical coherent states (CCS) given by 

\beq
 \langle N^2 \rangle_{\bf z}  = |{\bf z}|^4 + |{\bf z}|^2, \quad \quad  \langle N \rangle_{\bf z} = |{\bf z}|^2
\eeq

leading to the following Mandel parameter related to a  Poissonian statistics:

\beq
\mathcal Q_{\mbox{ccs}} &=& \frac{(|{\bf z}|^4 + |{\bf z}|^2) - (|{\bf z}|^2)^2}{|{\bf z}|^2} - 1 \cr
&=& 0.
\eeq

For the CS $|\eta_{\bf z}\rangle_q$, the Mandel parameter denoted by  $\mathcal Q_q$ is obtained as 

\beq
\mathcal Q_q &=& \frac{\left(\mathcal N_q(|{\bf z}|^{2})\right)^{-1}\sum_{ {\bf n}=0}^{\infty}
  \frac{|{\bf z}|^{2{\bf n}}}{[{\bf n}]_q !}  [{\bf n}]^2_q  - (|{\bf z}|^{2})^2}
{|{\bf z}|^{2}} -1 \cr
&=& \left(\mathcal N_q(|{\bf z}|^{2})\right)^{-1}\sum_{ {\bf n}=0}^{\infty}
  \frac{|{\bf z}|^{2{\bf n}}}{[{\bf n}]_q !}  [{\bf n}+1]_q  - (|{\bf z}|^{2} + 1) 
\eeq
such that in the limit $q\rightarrow 1$ we get $\mathcal Q_q  \rightarrow \mathcal Q_{\mbox{ccs}}=0$.

Note that in \cite{nge-hk}, the quantum statistical properties of the deformed  
states are   discussed   in the context of
conventional as well as deformed quantum optics.

Introducing  the $f$-deformed $SU_f(1,1)$ algebra (see \cite{liu-li} and references therein), which consists of three generators

\beq
\mathcal K_- = A_l A_k, \quad \mathcal K_+ = A'^{\dag}_l A'^{\dag}_k, \quad \mathcal K_0 = \frac{1}{2}(\mathcal N_l + \mathcal N_k + \mathbb I), 
\quad  \mathcal N_l = A'^{\dag}_l A_l \quad \mbox{and} \quad  \mathcal N_k = A'^{\dag}_k A_k \nonumber
\\
\eeq
where $A_l,  A_k$ and $A'^{\dag}_l,   A'^{\dag}_k\, (1 \leq k, l \leq N_B)$ are given by (\ref{oprealiz01}) and 
(\ref{oprealiz02}), we get the following  commutation relations
\beq
[\mathcal K_+, \mathcal K_-] = -2\mathcal K_0, \qquad [\mathcal K_0, \mathcal K_{\pm}] = \pm \mathcal K_{\pm}. 
\eeq

This  algebra is a generalization of the
$SU(1, 1)$ Lie algebra  \cite{perelomov}. Indeed, when  $\mathcal K_+$ and $\mathcal K_-$ are Hermitian conjugate to each other in 
the special case of $f(N_i) = 1$, i.e., $\mathcal K^{\dag}_- = \mathcal K^{\dag}_+ $, the $SU_f(1, 1)$ algebra contracts to the
$SU(1, 1)$ Lie algebra. 

Let now
  
\beq
X = \frac{\mathcal K^{\dag}_- + \mathcal K_-}{2}, \qquad  Y = \frac{i(\mathcal K^{\dag}_- - \mathcal K_-)}{2}
\eeq 
be the $f$-deformed quadrature operators  satisfying the   commutation relation 

\beq\label{sucomm}
[X, Y] = \frac{i}{2} [\mathcal K_-, \mathcal K^{\dag}_-] = \frac{i}{2}[(N_k+1)f^2_k(N_k+1)(N_l+1)f^2_l(N_l+1) - N_kf^2_k(N_k) N_lf^2_l(N_l)].
\nonumber\\
\eeq
The following uncertainty relation 

\beq
\langle (\Delta X)^2 \rangle \langle (\Delta Y)^2 \rangle \geq \frac{1}{16}|\langle [\mathcal K_-, \mathcal K^{\dag}_-] \rangle|^2 
\eeq
holds  
%{\color{blue} 
and the   $SU_f(1, 1)$ squeezing is  provided by the relation:

\beq
  \langle (\Delta X_k)^2 \rangle   < \frac{1}{4}|\langle [\mathcal K_-, \mathcal K^{\dag}_-] \rangle|, \qquad X_k= X, Y.
\eeq

%} 

\section{Concluding remarks}
In this work we have provided a  construction  of a  dual pair of nonlinear coherent states (NCS) in the context of 
changes of bases in the underlying Hilbert space for a model Hamiltonian describing the electron-phonon dynamics in condensed 
matter physics, which obeys a $f$-deformed Heisenberg algebra. 
The existence and properties of reproducing kernel  in the NCS Hilbert space have been studied and discussed;    the  probability density
and its dynamics  in the  basis of constructed coherent states have been analyzed.   
A  Glauber-Sudarshan $P$-representation of the density matrix   and  relevant issues  related to the reproducing kernel 
 properties have been presented in both
the NCS and nonlinear VCS. Moreover,  a  NCS  quantization 
 of classical phase space observables has been performed and illustrated in the case  of the Quesne's $q$-deformed CS
corresponding to the 
situation where  the deformation structure function $f(N) = \sqrt{\frac{[N]_q}{N}}.$ Finally, 
 some significant quantum optical  properties such as  the SNR and the Mandel parameter have been inspected.

\section*{Acknowledgements}
This work is partially supported by the Abdus Salam International
Centre for Theoretical Physics (ICTP, Trieste, Italy) through the
Office of External Activities (OEA) - \mbox{Prj-15}. The ICMPA
is in partnership with
the Daniel Iagolnitzer Foundation (DIF), France.

\section*{Appendix A. Proof of the Proposition \ref{kernelprop} }
\label{app1}

\bitem

\item [(i)] Hermiticity  

From (\ref{kern01}), we get

\beq
\mathcal K({\bf z'}, {\bf z}) = \langle \eta_{\bf z}|\eta_{\bf z'} \rangle  =  
\frac{\mathcal N(\bar{\bf z}{\bf z'})}{\sqrt{\mathcal N(|{\bf z'}|^{2})\mathcal N(|{\bf z}|^{2})}}
\eeq

and thereby 

\beq
\overline{\mathcal K({\bf z'}, {\bf z})} &=& \overline{\langle \eta_{\bf z}|\eta_{\bf z'} \rangle }\cr 
&=& \frac{\mathcal N(\bar{\bf z'}{\bf z})}{\sqrt{\mathcal N(|{\bf z'}|^{2})\mathcal N(|{\bf z}|^{2})}}
= \mathcal K({\bf z}, {\bf z'}).
\eeq 

\item [(ii)] Positivity

Using again (\ref{kern01}), we get

\beq
\mathcal K({\bf z}, {\bf z}) = \langle \eta_{\bf z}|\eta_{\bf z} \rangle  &=&  
\frac{\mathcal N(\bar{\bf z}{\bf z})}{\sqrt{\mathcal N(|{\bf z}|^{2})\mathcal N(|{\bf z}|^{2})}}\cr
&=& \frac{\mathcal N(|{\bf z}|^2)}{\mathcal N(|{\bf z}|^{2})} = 1 > 0
\eeq

implying that $\mathcal K({\bf z}, {\bf z})  >  0.$

\item [(iii)]  Idempotence

The left-hand side of the expression (\ref{kern02}) can be written,  by use of the definition (\ref{kern01}),  as 

\beq
\int_{\mathcal D}  d\mu ({\bf z''},{\bar{\bf z''}}) \mathcal K({\bf z}, {\bf z''})\mathcal K({\bf z''}, {\bf z'}) &=& 
\int_{\mathcal D}  d\mu ({\bf z''},{\bar{\bf z''}})
\left\{\frac{\mathcal N(\bar{\bf z''}{\bf z})}{\sqrt{\mathcal N(|{\bf z}|^{2})\mathcal N(|{\bf z''}|^{2})}}\right\} \cr
&& \times \left\{\frac{\mathcal N(\bar{\bf z'}{\bf z''})}{\sqrt{\mathcal N(|{\bf z'}|^{2})\mathcal N(|{\bf z''}|^{2})}}\right\}\cr
&=& \left\{\frac{1}{\sqrt{\mathcal N(|{\bf z}|^{2})\mathcal N(|{\bf z'}|^{2})}}
\sum_{{\bf n}=0}^{\infty}
\frac{{\bar{\bf z'}}^{\bf n} {\bf z}^{\bf n} }{\left\{{\bf n}\right\}  !}
\right\} \cr
&& \times \int_{\mathcal D}  d\mu ({\bf z''},{\bar{\bf z''}}) \frac{1}{\mathcal N(|{\bf z''}|^{2})}\left\{\sum_{{\bf n}=0}^{\infty}
\frac{|{\bf z''}|^{2}}{\left\{{\bf n}\right\}  !}
\right\}\cr
\int_{\mathcal D}  d\mu ({\bf z''},{\bar{\bf z''}}) \mathcal K({\bf z}, {\bf z''})\mathcal K({\bf z''}, {\bf z'})
&=&  \frac{1}{\sqrt{\mathcal N(|{\bf z}|^{2})\mathcal N(|{\bf z'}|^{2})}}
\sum_{{\bf n}=0}^{\infty}
\frac{{\bar{\bf z'}}^{\bf n} {\bf z}^{\bf n} }{\left\{{\bf n}\right\}  !} 
\cr
&=:&  \mathcal K({\bf z}, {\bf z'})
\eeq
where the following relations 

\beq
\frac{1}{\mathcal N(|{\bf z}|^{2})}\sum_{{\bf n}=0}^{\infty} 
\frac{|{\bf z}|^{2}}{\left\{{\bf n}\right\}  !} 
 = 1, \qquad \int_{\mathcal D}  d\mu ({\bf z},{\bar{\bf z}}) = 1
\eeq

are used.

\eitem

$\hfill{\square}$

\section*{Appendix B. Proof  of the  Proposition \ref{densmatkernel}}
\label{app2}

\bitem

\item [(i)] Hermiticity  
 
We have 
\beq
\rho_{[{\bf k}]}({\bf z}, {\bf z'}) = \langle \eta_{\bf z}   |\rho_{[{\bf k}]}|\eta_{\bf z'} \rangle
\eeq
 
and thereby 

\beq
\overline{\rho_{[{\bf k}]}({\bf z}, {\bf z'})} &=& \overline{\langle \eta_{\bf z}   |\rho_{[{\bf k}]}|\eta_{\bf z'} \rangle} \cr
&=&  
\langle \eta_{\bf z'}   |\rho_{[{\bf k}]}|\eta_{\bf z} \rangle = \rho_{[{\bf k}]}({\bf z'}, {\bf z}).
\eeq

\item [(ii)] Positivity

%{\color{blue}
From (\ref{proba}) and (\ref{densmat07}), we get

\beq
\langle \eta_{\bf z} |\rho_{[{\bf k}]}|\eta_{\bf z} \rangle &=& 
\langle \eta_{\bf z} |\rho_{[{\bf k}]}
\int_{\mathcal D}  d\mu ({\bf z},{\bar{\bf z}}) P(|{\bf z}|^{2}) \,|\eta_{\bf z} \rangle \cr
&=& \sum_{{\bf m}, {\bf n} =0}^{\infty}\int_{\mathcal D}d\mu ({\bf z},{\bar{\bf z}})\, \left(\mathcal N(|{\bf z}|^{2})\right)^{-1} 
 P(|{\bf z}|^{2}) \,
\rho_{[{\bf k}]}({\bf n}, {\bf m}) \frac{{r}^{{\bf m}+{\bf n}} 
e^{-\imath ({\bf n}-{\bf m})\theta} }{\sqrt{\left\{{\bf n}\right\}  !}\sqrt{\left\{{\bf m}\right\}  !}}\cr
&=&   \sum_{{\bf n} =0}^{\infty} \rho_{[{\bf k}]}({\bf n}, {\bf n}) \,
\left\{\int_{\mathcal D} d\mu ({\bf z},{\bar{\bf z}})  P(|{\bf z}|^{2}) \left(\mathcal N(|{\bf z}|^{2})\right)^{-1} 
\frac{|{\bf z}|^{2\bf n}}{\left\{{\bf n}\right\}  !} \right\}  \cr
&=&   \sum_{{\bf n} =0}^{\infty} [\rho_{[{\bf k}]}({\bf n}, {\bf n})]^2 \cr
&=& \sum_{{\bf n} =0}^{\infty} \left\{2\pi \int_{0}^{L}
\frac{{r}^{2{\bf n} }}{\left\{{\bf n}\right\}  !} \frac{P(r^2)}{\mathcal N(r^{2})} d\lambda(r) \right\}^2   > 0
\eeq

implying that

\beq
\langle \eta_{\bf z} |\rho_{[{\bf k}]}|\eta_{\bf z} \rangle = \rho_{[{\bf k}]}({\bf z}, {\bf z})  >  0.
\eeq
%}

\item [(iii)]  Idempotence

By  setting $d\mu ({\bf z''},{\bar{\bf z''}}) = \mathcal N(|{\bf z''}|^{2}) d\lambda(r'') d\theta$ and using the relation 
$\sum_{{\bf n} =0}^{\infty} \rho_{[{\bf k}]}({\bf n}, {\bf n})   = 1$, we get from (\ref{densmat05})

\beq
&&\int_{\mathcal D}  d\mu ({\bf z''},{\bar{\bf z''}}) \rho_{[{\bf k}]}({\bf z}, {\bf z''})\rho_{[{\bf k}]}({\bf z''}, {\bf z'})\cr
&&=
\int_{\mathcal D}  d\mu ({\bf z''},{\bar{\bf z''}})\left[\left[\mathcal N(|{\bf z}|^{2}) \mathcal N(|{\bf z''}|^{2})\right]^{-1/2}
\sum_{{\bf n}, {\bf m} =0}^{\infty} \rho_{[{\bf k}]}({\bf n}, {\bf m})
\frac{\bar{\bf z}^{\bf n}}{\sqrt{\left\{{\bf n}\right\}  !}}
\frac{{\bf z''}^{\bf m}}{\sqrt{\left\{{\bf m}\right\}  !}} \right] \cr
&&\times \left[\left[\mathcal N(|{\bf z''}|^{2}) \mathcal N(|{\bf z'}|^{2})\right]^{-1/2}
\sum_{{\bf n}, {\bf m} =0}^{\infty} \rho_{[{\bf k}]}({\bf n}, {\bf m})
\frac{\bar{\bf z''}^{\bf n}}{\sqrt{\left\{{\bf n}\right\}  !}}
\frac{{\bf z'}^{\bf m}}{\sqrt{\left\{{\bf m}\right\}  !}} \right]\cr 
&&=
\left[\left[\mathcal N(|{\bf z}|^{2}) \mathcal N(|{\bf z'}|^{2})\right]^{-1/2}
\sum_{{\bf n}, {\bf m}  =0}^{\infty} \rho_{[{\bf k}]}({\bf n}, {\bf m})
\frac{\bar{\bf z}^{\bf n}}{\sqrt{\left\{{\bf n}\right\}  !}}
\frac{{\bf z'}^{\bf m}}{\sqrt{\left\{{\bf m}\right\}  !}} \right] \cr
&&\times 
\left[ \sum_{{\bf n} =0}^{\infty} \rho_{[{\bf k}]}({\bf n}, {\bf n})  
\right] \cr
&&= \rho_{[{\bf k}]}({\bf z}, {\bf z'}).
\eeq

\eitem 

$\hfill{\square}$

\end{document}